\documentclass[oldversion]{aa}
\usepackage{graphicx}
\usepackage{array}
\usepackage{tabularx}
\usepackage{natbib}
\setcitestyle{aysep={}} 
\usepackage{hyperref}
\usepackage{longtable}
\usepackage{lscape}
\usepackage{siunitx}
\usepackage{caption}
\usepackage{subcaption}
\usepackage{color,ulem}
\usepackage{nicefrac}
\newcommand{\teff}{$\mathrm{T_{eff}}$}
\newcommand{\av}{$\mathrm{A_V}$}

\newcommand{\masyr}{mas\,yr$^{-1}$}

\newcommand{\new}{\textcolor{blue}}
\begin{document}

   \title{BEAST detection of a brown dwarf and a low-mass stellar companion around the young bright B star HIP 81208
\thanks{Based on observations from the European Southern Observatory, Chile (Programme 1101.C-0258).}
}


   \author{Gayathri Viswanath\inst{1} \and
          Markus Janson\inst{1} \and
          Raffaele Gratton\inst{2} \and
          Vito Squicciarini\inst{3,2} \and
          Laetitia Rodet\inst{4} \and
          Simon C. Ringqvist\inst{1} \and
          Eric E. Mamajek\inst{5,6} \and
          Sabine Reffert\inst{7} \and
          Ga{\"e}l Chauvin\inst{8} \and
          Philippe Delorme\inst{9} \and
          Arthur Vigan\inst{10} \and
          Micka{\"e}l Bonnefoy\inst{9} \and
          Natalia Engler\inst{11} \and
          Silvano Desidera\inst{2} \and
          Thomas Henning\inst{12} \and
          Janis Hagelberg\inst{13} \and
          Maud Langlois\inst{14} \and
          Michael Meyer\inst{15}
          }

         \institute{Institutionen f{\"o}r astronomi, Stockholms Universitet, Stockholm, Sweden\\
              \email{gayathri.viswanath@astro.su.se}
        \and
           INAF - Osservatorio Astronomico di Padova, Padova, Italy
         \and
            LESIA, Observatoire de Paris, Université PSL, CNRS, Sorbonne Université, Université de Paris, 5 place Jules Janssen, 92195 Meudon, France
        \and
           Cornell Center for Astrophysics and Planetary Science, Department of Astronomy, Cornell University, Ithaca, NY, USA
        \and
           Jet Propulsion Laboratory, California Institute of Technology, Pasadena, CA, USA
         \and
           Department of Physics and Astronomy, University of Rochester, Rochester, NY, USA
        \and
           Landessternwarte, Zentrum f{\"u}r Astronomie der Universit{\"a}t Heidelberg, Heidelberg, Germany
        \and
            Laboratoire J.-L. Lagrange, Universit{\'e} Cote d’Azur, CNRS, Observatoire de la Cote d’Azur, 06304 Nice, France
        \and
           Univ. Grenoble Alpes, IPAG, Grenoble, France
         \and
           Aix Marseille Universit{\'e}, CNRS, LAM, Marseille, France
        \and
            ETH Zurich, Zurich, Switzerland
        \and
            Max Planck Institut f{\"u}r Astronomie, Heidelberg, Germany
        \and
            Observatoire de Geneve, University of Geneva,
            Ch. de Pegasi 51, 1290 Versoix, Switzerland
         \and
            CRAL, CNRS, Universit{\'e} Lyon, Saint Genis Laval, France
        \and
            Department of Astronomy, University of Michigan, Ann Arbor, MI, USA
             }

   \date{Received ---; accepted ---}

   \abstract{
   Recent observations from B-star Exoplanet Abundance Study (BEAST) have illustrated the existence of sub-stellar companions around very massive stars. In this paper, we present the detection of two lower mass companions to a relatively nearby ($148.7^{+1.5}_{-1.3}$ pc), young ($17^{+3}_{-4}$ Myr), bright (V=$6.632\pm0.006$ mag), $2.58\pm0.06~ M_{\odot}$ B9V star HIP 81208 residing in the  Sco-Cen association, using the Spectro-Polarimetric High-contrast Exoplanet REsearch (SPHERE) instrument at the Very Large Telescope (VLT) in Chile. Analysis of the photometry obtained gives mass estimates of $67^{+6}_{-7}~M_J$ for the inner companion and $0.135^{+0.010}_{-0.013}~M_{\odot}$ for the outer companion, indicating the former to be most likely a brown dwarf and the latter to be a low-mass star. The system is compact but unusual, as the orbital planes of the two companions are likely close to orthogonal. The preliminary orbital solutions we derived for the system indicate that the star and the two companions are likely in a Kozai resonance, rendering the system dynamically very interesting for future studies.
   }


\titlerunning{Detection of HIP 81208 B and C}
\authorrunning{G. Viswanath et al.}

   \maketitle
%

\section{Introduction}
\label{s:intro}

Most known exoplanets are detected through the transit and radial velocity techniques, due to their exquisite sensitivity to planets in small orbits around their host stars \citep[e.g.][]{mayor1995,winn2011}. However, both techniques are much less sensitive to planets in wide orbits (several au or larger), and therefore this class of planets has been much less explored. In recent years, a growing population of wide planets have been discovered with direct imaging \citep[e.g.][]{macintosh2015,chauvin2017,keppler2018,bohn2020} to an increasing extent in synergy with astrometry \citep[e.g.][]{brandt2021,lacour2021, hinkley2022, franson2023}. These techniques are also sensitive to brown dwarf companions \citep[e.g.][]{janson2012,crepp2016,currie2020}; while such companions are rare, their high masses and brightnesses make them suitable for detailed characterisation and forming important links for understanding atmospheric conditions down to progressively smaller substellar masses and temperatures.

Direct imaging surveys \citep[e.g.][]{uyama2017, nielsen2019, ispy2020, vigan2021} have studied wide planet and brown dwarf companion demographics in a range of stellar system types, including binaries \citep[e.g.][]{bonavita2016,asensio2018}, debris disk systems \citep[e.g.][]{janson2013,meshkat2017} and stars of different masses \citep[e.g.][]{bowler2012,delorme2012,wagner2022}. Studying substellar companion demographics as function of stellar mass can yield insights into the planet and brown dwarf formation processes, and in this context, the most massive stars in the Solar neighbourhood -- B stars -- represent an important part of the puzzle. To probe this regime deeply for the first time, we have initiated the B-star Exoplanet Abundance Study \citep[BEAST, see][]{janson2021a}, which is a high-contrast imaging survey using the Spectro-Polarimetric High-contrast Exoplanet REsearch \citep[SPHERE; ][]{beuzit2019} Extreme Adaptive Optics instrument at the Very Large Telescope (VLT) in Chile, to study 85 B stars in the nearby young Sco-Cen region \citep{dezeeuw1999}. Currently ongoing, BEAST has already discovered several new planets and low-mass substellar companions \citep{janson2019,janson2021b,squicciarini2022a}.

Here, we report on the discovery of two low-mass companions, both in a dynamically unusual orbital configuration around the B9 star HIP 81208. The paper is structured as follows: in Section \ref{s:obs}, we give the details of the observation of the target and subsequent data reduction. In Section \ref{s:star}, we describe the target star and derive some of its basic parameters. Section \ref{s:results} describes the main results from this work, including characterisation of the candidate properties, a detailed analysis of the spectrum of the inner companion candidate, and analysis of the orbits of the two companion candidates, as well as a discussion on the dynamical stability of the system. The main conclusions from this study are outlined in Section \ref{s:summary}.

\section{Observations and data reduction}
\label{s:obs}

The target system, HIP 81208, has been observed twice with the SPHERE instrument at the Very Large Telecsope (VLT), both in the context of BEAST. The first observation was acquired on 6 Aug 2019 (MJD 58701.02), and a second-epoch follow-up observation was acquired on 5 Apr 2022 (MJD 59674.35). Both observations used the standard settings for BEAST \citep{janson2021a}, in the IRDIFS-EXT mode \citep{zurlo2014} that allows for simultaneous low-resolution $YJH$-band spectroscopy using the Integral Field Spectrograph \citep[IFS; ][]{beuzit2019} and dual-band imaging in the $K$ band with the $K12$ filter-pair \citep{vigan2010} using the Infra-Red Dual-band Imager and Spectrograph \citep[IRDIS; ][]{beuzit2019}. 

The main observational sequence was executed in pupil-stabilised mode, which facilitates Angular Differential Imaging \citep[ADI, see][]{marois2006} reductions, and utilised the N-ALC-YJH-S coronagraph to enhance the instrumental contrast close to the parent star. This long coronagraphic sequence was enveloped by two short pairs of images: A short unsaturated (non-coronagraphic) exposure of the star was acquired before and after the sequence for spectrophotometric calibration purposes, and a short coronagraphic exposure was taken with the so-called waffle mode \citep{cantalloube2019} turned on, which generates satellite images of the primary star for astrometric calibration purposes. An empty sky frame was also included in the observational procedure for the purpose of sky subtraction. The long coronagraphic sequence consisted of 16 frames, with 3 sub-integrations of 64 seconds each. Hence, the total integration time on-target was 3072 seconds in both epochs. The total field rotations during the observations were 51.9 deg in the 2019 epoch and 57.5 deg in the 2022 epoch.

Data reduction was performed according to the regular BEAST procedure \citep{janson2021a}, with the SPHERE Data Center\new{\footnote{\url{https://sphere.osug.fr/spip.php?article45}} }\citep{delorme2017} software based on the SPHERE pipeline \citep{pavlov2008} for all basic calibration steps including dark and sky subtraction, flat fielding, and spectral extraction. For pixel scale, true North orientation and pupil offset, we used the long-term values given in \cite{maire2021} which are based on a five year analysis of the astrometric stability of SPHERE relying on observations of star clusters taken at regular intervals. Accordingly, for both epochs, the pixel scale adopted for IRDIS was 12.258$\pm$0.004 mas/pixel for $K1$ band and 12.253$\pm$0.003 mas/pixel in the $K2$ band, while the pixel scale for IFS was $7.46\pm0.02$ mas/pixel. The true North orientation used was -1.77$\pm$0.04 deg and the pupil offset was 136$\pm$0.03 deg. For ADI reduction purposes, we used SpeCal \citep{galicher2018} for IRDIS in Template Locally Optimised Combination of Images \citep[TLOCI; ][]{marois2014} mode, and a pipeline for IFS based on Karhunen-Lo{\`e}ve Image Projection \citep[KLIP, see e.g.][]{soummer2012}. 

\section{Stellar properties}
\label{s:star}
HIP 81208 (alias HD 149274, TYC 7357-207-1, TIC 280474618, Gaia DR3 6020514769906985728, 2MASS J16351384-3543287) is a young bright (V=$6.632\pm0.006$ mag) B9V star \citep{houk1982} residing in the Upper Centaurus Lupus (UCL) sub-region of the Sco-Cen stellar association \citep{hoog2000}, at a distance of $148.7^{+1.5}_{-1.3}$ pc from Earth. Some basic measurements of the star obtained from the literature are listed in Table \ref{tab1}.

\renewcommand{\arraystretch}{1.2}
\begin{table*}[!ht]
\tiny
\centering
\resizebox{0.9\textwidth}{!}{%
\begin{tabular}[l]{l|l|l}
\hline\hline
\textbf{Stellar Parameter} & \textbf{Value}         & \textbf{Reference} \\\hline
ICRS coordinates (Ep=2016.0) & $\alpha = 248.80761058629^{\circ}$, $\delta = -35.72476117010^{\circ}$  & \cite{brown2022}  \\
Distance [pc]  & $148.7^{+1.5}_{-1.3}$ & \cite{brown2022} \\
Parallax [mas] & $6.842 \pm 0.048$ & \cite{brown2022} \\
$[\nicefrac{M}{H}]$ [dex] & $-0.217^{+0.217}_{-0.184}$  & \cite{Anders2022}\\
$\mu_{\alpha^*}$ [\masyr] & $-9.701\pm 0.052$ & \cite{brown2022} \\
$\mu_{\delta}$ [\masyr] & $-25.913\pm 0.039$ & \cite{brown2022} \\
V [mag] & $6.632\pm0.006$ & \cite{slawson1992} \\
B$-$V [mag] & $-0.049\pm0.007$ & \cite{esa1997} \\
U$-$B [mag] & $-0.208\pm0.009$ & \cite{slawson1992}\\
G [mag] & $6.6297\pm0.0028$ & \cite{brown2022} \\
J [mag] & $6.731\pm0.026$  & \cite{cutri2003} \\
H [mag] & $6.773\pm0.051$ & \cite{cutri2003} \\
Ks [mag] & $6.768\pm0.029$ & \cite{cutri2003} \\\hline
\end{tabular}%
}
\caption{Basic stellar parameters for HIP 81208 from the literature}
\label{tab1}
\end{table*}

The current literature gives no indication of a binary companion to HIP 81208. A potentially interesting Gaia star in the vicinity of HIP 81208 is discussed in Section \ref{sec:potential_D}. The proper motion anomaly (PMa) between the long-term Hipparcos-Gaia DR3 and short-term Gaia DR3 proper motion vectors is very low \citep[PMaRA = 0.063$\pm$0.056 \masyr, PMaDE = 0.090$\pm$0.041 \masyr; see][]{kervella2022}, ruling out the existence of any ($>0.05~M_{\odot}$) binary companion to the star within 3--30 au. There is also no hint of a faint or unresolved astrometric binary companion within 2--20 au of the star, in Gaia DR2, as inferred from the low value of renormalized unit weight error (RUWE), $\rho=0.898$ \citep{Belokurov2020}. Additionally, \cite{stock2021} analysed the 26 XSHOOTER \citep{vernet2011} spectra available in the ESO archive and derived radial velocities with a precision of around 2 kms$^{-1}$. The radial velocities cover a period of slightly less than 8 years from 2010 to 2017 and have an rms of about 5 kms$^{-1}$. The analysis revealed no indications of periodic variations, except for a slightly significant period of one year which also shows up in the window function. From this, we conclude that there are no indications of spectroscopic companions to HIP 81208 in the XSHOOTER radial velocities. 

In the following subsections, we constrain the main parameters of the target, a summary of which is also listed in Table \ref{tab1b}. 

\begin{figure*}[htbp!]
\centering
    \includegraphics[width=0.9\linewidth]{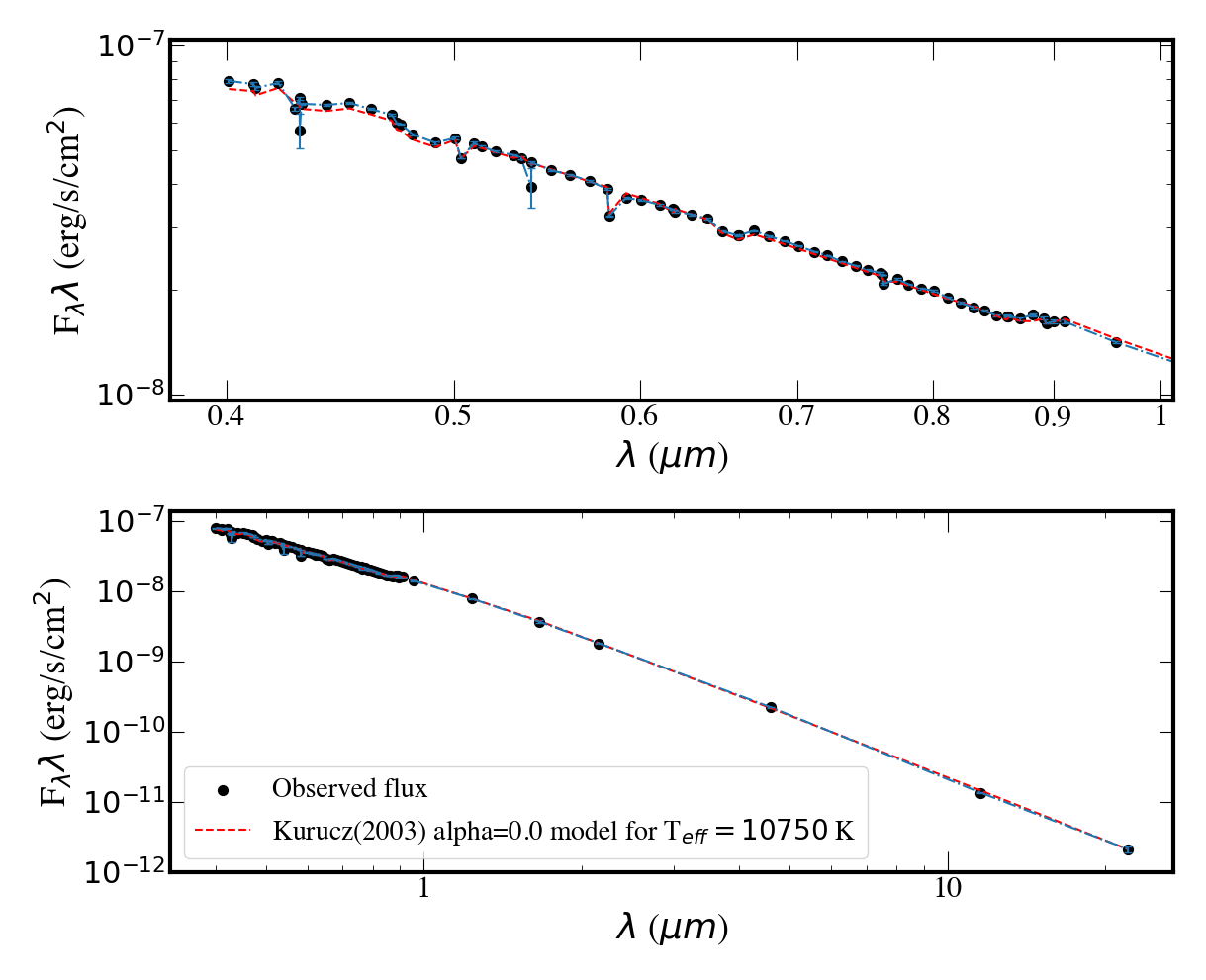}
    \caption{Spectral energy distribution of HIP 81208 obtained with 78 existing photometry points retrieved from the VO SED Analyzer (VOSA). Also shown is the best-fit stellar model \citep[Kurucz ODFNEW /NOVER, alpha: 0.0 (2003);][]{Castelli2003} for a \teff=10750 K. }
    \label{fig1}
\end{figure*} 

\subsection{Extinction and reddening} \label{extinction}
GaiaDR3 lists a monochromatic extinction estimate of $\mathrm{A_0}=0.0229^{+0.0035}_{-0.0034}$ mag for the target at $\lambda_0=541.4$ nm \citep[see][for the definition of $A_0$]{creevey2022}. \av, the extinction in the Johnson V band centred at $\sim$548 nm, is dependent on the spectrum of the emitting source in the V band and hence is intrinsically different from $A_0$, which is a property of the interstellar medium alone and is affected only by the amount of absorption there-in. However, this dependence of \av~on the source spectrum is often negligible, especially for small extinctions ($A_0<$ 2 mag), at which $A_0$ and \av~follow nearly an identity relation (see Fig 11.4(a) in the Gaia DR3 online documentation\footnote{\url{https://gea.esac.esa.int/archive/documentation/GDR3/Data_analysis/chap_cu8par/sec_cu8par_data/ssec_cu8par_data_xp.html##Ch11.F4}}). Hence, in this case, $A_0$ can be taken as a rough approximation of \av~for our target.
The STILISM reddening maps\footnote{\url{https://stilism.obspm.fr/}} from \cite{lallement2018} at the position and distance of HIP 81208 gives a reddening estimate E(B$-$V)=$0.011\pm0.021$ mag, which can be translated to an \av~value of $0.034\pm0.064$ mag using the relation \av=3.07$\times$E(B$-$V) \citep[][appropriate for lightly reddened A0 V stars]{Mccall2004}. Examination of the reddening values for stars within $\pm4^{\circ}$ of the position of HIP 81208 in the \cite{Reis2011} catalog  shows that stars in its vicinity are essentially negligibly reddened until a distance of $\sim$160 pc, and the stars in the distance range 170--180 pc have \av=0.1--0.2 mag. 
In particular, four stars (HD 147493, HD147387, HD147149, HD147597) in the immediate vicinity of HIP 81208, with distances 146--150 pc, have negligible reddening values of E(B$-$V)=$0.000\pm0.008$, $-0.008\pm0.011$, $0.003\pm0.012$ and $0.009\pm0.011$ mag respectively, with an inferred \av=$0.00\pm0.03$, $-0.03\pm0.05$, $0.01\pm0.05$ and $0.04\pm0.05$ mag respectively. Given the similar distance and similar position of these stars in the sky as our target, one can expect them to have comparable reddening and extinction as HIP 81208.

An independent estimate of this B star's reddening and extinction can be obtained from UBV photometry using the Q-method to calculate the reddening-free Q index \citep[e.g.][see Appendix C.3 of \cite{PM2013}]{JM1953}. Adopting the U$-$B colour from \cite{slawson1992} and the B$-$V colour from \cite{esa1997}, and the updated calibration of the Q-index to the dereddened (B$-$V)$_0$ and (U$-$B)$_0$ colours from \cite{PM2013}, we estimate the intrinsic colours to be (B$-$V)$_0$=$-0.075\pm0.003$ mag and (U$-$B)$_0$=$-0.228\pm0.013$ mag, with reddening E(B$-$V)=$0.026\pm0.009$ mag and extinction \av=$0.085\pm0.030$ mag. 
Additionally, a fit of the \cite{PM2013} MS colour grid to the UBVJHKs photometry for HIP 81208, allowing the E(B$-$V) to vary, finds that the reduced $\chi^2 < 1$ fits result in estimates of \teff=$10970\pm350$K, E(B$-$V)=$0.020\pm0.013$ mag and \av=$0.064\pm0.041$ mag for the target.

Both the reddening estimates derived above -- using the Q-method and the spectral energy distribution (SED) fitting -- rely heavily on the UBV photometry for a single star, i.e. HIP 81208, and hence are not independent. However, both these estimates (E(B$-$V)$\leq$0.02$\pm$0.02 mag) are essentially consistent with the STILISM reddening map from \cite{lallement2018} (E(B$-$V)=0.011$\pm$0.021 mag), which has the benefit of having already been published, and in addition is also a reddening estimate averaged over a number of stars of varying distances and positions in the vicinity of HIP 81208. The latter estimate is thus more robust than the other estimates of reddening derived/quoted in this Section, although the consistency between them adds to its confidence. We thus adopt the STILISM value of reddening and the inferred interstellar extinction thereafter,  \av=0.034$\pm$0.064 mag, for the target in this paper.

\subsection{Spectral energy distribution and effective temperature}
\label{sec:teff}

We used the Virtual Observatory SED analyzer (VOSA\footnote{\url{http://svo2.cab.inta-csic.es/theory/vosa/}}) version 7.5 to query for the photometry of HIP 81208 and fit synthetic stellar spectra to the data. The observed spectrum for 78 photometric data points is shown in Fig. \ref{fig1}. We removed several discrepant or redundant photometry points, namely the synthetic Gaia DR3 data (for e.g, $U, B, V, R, I$) as they were dependent on other available Gaia DR3 photometry, and APASS Sloan photometry. Gaia DR3 photometry from binned $Bp$ and $Rp$ spectra account for 66 of the data points, along with the Gaia DR3 $G_{BP}/G/G_{RP}$ points, and photometry from Tycho (B$T$, V$T$), APASS ($B, V$), 2MASS ($J,H,Ks$), WISE ($W1, W2, W3, W4$). For the SED fitting, we assumed a prior range of \av$ = 0.034 \pm 0.064$ mag. Using Kurucz 2003 [a/Fe]=0.0 models \citep{Castelli2003}, the best overall fit that fits all 78 data points had the following parameters: \av=$-0.03$ mag, \teff=$10750\pm125$ K, log(\textsl{g})=4, $[\nicefrac{M}{H}]$=$-1$, $\mathcal{F}_{bol}$=$(8.43\pm0.03)\times10^{-8}$ erg~cm$^{-2}$~s$^{-1}$, $\mathcal{L}_{bol}$=$58.27\pm1.304\,\mathcal{L}_{\odot}$ $(log(\mathcal{L}/\mathcal{L}_{\odot})$=$1.765\pm0.01)$.
The existing photometry as well as the obtained SED fit does not show any infrared excess up to $\sim10~\mu m$ (see lower panel of Fig. \ref{fig1}), which indicates the possible absence of any disk around HIP 81208. The SED fitting clearly favored models with negligible or zero reddening. The parameters all seem reasonable, except for the low metallicity, which was also found by the Gaia DR3 GSP Phot Aeneas fit to the Bp/Rp spectra ([$\nicefrac{M}{H}$]=$-1.27^{+0.39}_{-0.21}$). Such a low metallicity is unexpected as the metallicity of Sco-Cen is essentially solar, however the B9V star may be suffering from some abundance anomaly or other effect (e.g., fast rotation) which may be biasing the SED fitting.

We had already obtained a \teff=10970 K from the SED fit of MS colour grid to the UBVJHKs photometry in Section \ref{extinction}.
From the inferred colours from the Q-method described before, the latest version of the \cite{PM2013} table\footnote{\label{pmtable}\url{https://www.pas.rochester.edu/~emamajek/EEM_dwarf_UBVIJHK_colors_Teff.txt}} suggest that these values of intrinsic colours correspond to a \teff~of 11130$\pm$110 K (and indeed very typical for B9V dwarfs).

Furthermore, the star's spectral type itself (B9V) suggests a \teff~of approximately 10700 K, based on values for B9V standard stars by \cite{PM2013}.

An additional independent \teff~estimate can be derived from the H$_{\beta}$ photometry from \cite{slawson1992} ($\beta$=2.842$\pm$0.005 mag). Using the ATLAS9 model grid from \cite{Castelli2003} for assumed atmospheric parameters log(\textsl{g})=4.0, $[\nicefrac{M}{H}]$=0, and $v_{turb}=2\,km\,s^{-1}$, the Slawson H$_{\beta}$ value corresponds to a \teff=10654 K (note that adopting $v_{turb}$=0 km\,s$^{-1}$ has negligible effect, yielding \teff=10679 K).

Hence, we derived several \teff~estimates ranging from 10654 K to 11130 K. These five estimates are not quite fully independent as three of the estimates are connected to the \cite{PM2013} dwarf color/\teff~sequences, but they are consistent with an approximate \teff~of 10840$\pm$220 K (rms). This value is within 2.5$\sigma$ of the Gaia DR3 GSP-phot Aeneas estimate of $11396^{+57}_{-76}$ K.

\subsection{Luminosity and Radius}
We calculated a new bolometric luminosity for our target from the \av~ and \teff~ obtained in Section \ref{extinction} and Section \ref{sec:teff} and V mag from Table \ref{tab1}. We converted V to m$_{bol}$ using a bolometric correction of BC$_v=-0.447$ mag, obtained from interpolating the PM2013 table (refer footnote\ref{pmtable}) for a \teff=10840 K. The resulting apparent bolometric magnitude is m$_{bol}$=$6.154\pm0.064$ mag. At a distance of 148.7 pc, this corresponds to the absolute magnitude M$_{bol}$=$0.292\pm0.080$ mag. This translates to an equivalent luminosity of $\mathcal{L}_{bol}$=$60.69\pm4.46$ $\mathcal{L}_{\odot}$ $(log(\mathcal{L}/\mathcal{L}_{\odot})$=$1.783\pm0.032$).
This is consistent with previous $\mathcal{L}_{bol}$ values for HIP 81208 in the literature \citep[e.g.,][]{xhip}. 

Using the above derived value for $\mathcal{L}_{bol}$ and the \teff~estimate from Section \ref{sec:teff} ($10840\pm220$ K), we can arrive at an estimate for the radius of the target using the Stefan-Boltzmann law, as $\mathcal{R}_{*}=\sqrt{\nicefrac{\mathcal{L}_{bol}}{4\pi\sigma T_{eff}^4}}=2.213\pm0.121~R_{\odot}$.

\subsection{Stellar age}
The stellar age was constrained using the method of isochronal dating of co-moving stars (CMS) introduced in the previous BEAST publications \cite{janson2021a} and \cite{squicciarini2022a}. We adopted the long-term proper motion for HIP 81208 reconstructed by \cite{kervella2022}, for this purpose. The sample of co-moving stars was constructed by querying the Gaia DR3 catalogue within a search radius of 5 deg around the location of our target, for sources possessing similar 2D-projected space motion. A clump of 86 such sources was identified in the velocity space ($v_\alpha$, $v_\delta$) through the following cuts on velocity and parallax ($\varpi$):
\begin{equation} \label{eq:coord_transf1}
    \begin{cases}
    -7.5 \text{ km s}^{-1} < v_\alpha < -5.8 \text{ km s}^{-1} \\
    -19.5 \text{ km s}^{-1} < v_\delta < -15.5 \text{ km s}^{-1} \\
    6.0 \text{ mas} < \varpi < 7.3 \text{ mas}
\end{cases}
\end{equation}
Based on their photometry from Gaia DR3 ($G$, $G_{BP}$, $G_{RP}$) and 2MASS ($J$, $H$, $K_s$) bands, we subsequently derived the isochronal ages and masses for all the stars in the group using the \textsc{madys} tool \citep{squicciarini2022b}. Three independent evolutionary models, all assuming solar metallicities, were employed to ensure the robustness of the results: BHAC15 \citep{baraffe2015}, PARSEC \citep{marigo2017} and MIST \citep{choi2016}. The sample was then divided into 5 mass bins: $\nicefrac{M}{M_{\odot}}$$\leq$0.4, 0.4<$\nicefrac{M}{M_{\odot}}$$\leq$0.6, 0.6<$\nicefrac{M}{M_{\odot}}$$\leq$0.8, 0.8<$\nicefrac{M}{M_{\odot}}$$\leq$1.4, $\nicefrac{M}{M_{\odot}}$>1.4. We restricted our age analysis to the most reliable bin of stellar mass ($0.8 M_\odot<M\leq1.4 M_\odot$) containing 9 out of the 86 stars in the sample. The small number of stars in this bin is a consequence of the shape of the initial mass function (IMF) (53 of the fitted stars were M-type stars, i.e. M<0.6 $M_{\odot}$) and the fact that not all the stars could be successfully fitted due to factors like non-optimal quality of photometric data or unresolved binarity. Using the chosen mass bin, we were able to derive a well-defined group age estimate $t=17_{-4}^{+3}$ Myr, which is consistent with previous age estimates for UCL like \citet[][17$\pm$1 Myr]{Mamajek2002}, \citet[][16$\pm$1 Myr]{PM2012}.

\renewcommand{\arraystretch}{1.5}
\begin{table*}[!ht]
\tiny
\centering
\resizebox{0.9\columnwidth}{!}{%
\begin{tabular}[l]{l|l}
\hline\hline
\textbf{Stellar Parameter} & \textbf{Estimated Value}  \\\hline
E(B-V) [mag]                     & $0.011\pm0.021$            \\
Av [mag]                        & $0.034\pm0.064$   \\
T$_\mathrm{eff}$ [K]      & $10840\pm220$   \\
Age [Myr]     & $17^{+3}_{-4}$    \\
$\mathcal{L}_{bol}\,[\mathcal{L}_{\odot}]$  &  $60.69\pm4.46$ \\ 
$\mathcal{R}_*\,[\mathcal{R}_{\odot}]$    & $2.213\pm0.121$  \\
$\mathcal{M}_*\,[\mathcal{M}_{\odot}]$     & $2.58\pm0.06$\\
log(\textsl{g}) [cm/s$^2$] & $4.201\pm0.011$\\\hline
\end{tabular}%
}
\caption{A summary of all the stellar parameters estimated for HIP 81208 in this work.}
\label{tab1b}
\end{table*}

\subsection{Stellar mass and log({\textsl{g}})}
To estimate the stellar mass, we used the Parsec tracks \citep{bressan2012} version 1.2S \citep{chen2015} with A$_V$=0.034 mag and metallicity [$\nicefrac{M}{H}$] = 0 and age = 17 Myr. We interpolated for \teff =10840$\pm$220 K and log($\mathcal{L}/\mathcal{L}_{\odot}$)=1.7831$\pm$0.0319 to get an estimate for the mass of the star, $\mathcal{M}_{*}$ = 2.58$\pm$0.06 $\mathcal{M}_{\odot}$ and its surface gravity log(\textsl{g}) = 4.201$\pm$0.011. Uncertainties for $\mathcal{M}_{*}$ and log(\textsl{g}) were derived using the upper and lower limits for \teff ($\pm220$ K) and log($\mathcal{L}/\mathcal{L}_{\odot}$) ($\pm0.0319$). To get the upper limit on the mass and surface gravity, we interpolated the Parsec tracks for $\mathcal{M}_{*}$ and log(\textsl{g}) corresponding to \teff = 11060 K and log($\mathcal{L}/\mathcal{L}_{\odot}$) = 1.815. Similarly, to get the lower limits, we interpolated using \teff = 10620 K and log($\mathcal{L}/\mathcal{L}_{\odot}$) = 1.7512. The average of the uncertainties derived from the respective upper and lower limits of $\mathcal{M}_{*}$ and log(\textsl{g}) obtained thus are quoted as the uncertainty for these parameters in this Section. The values for $\mathcal{M}_{*}$ and log(\textsl{g}) derived here are also consistent with the literature \citep[see e.g.,][]{Anders2022}.

\section{Results and discussion}
\label{s:results}
\begin{figure*}
\centering
\begin{subfigure}{.5\textwidth}
  \centering
  \includegraphics[width=0.96\linewidth]{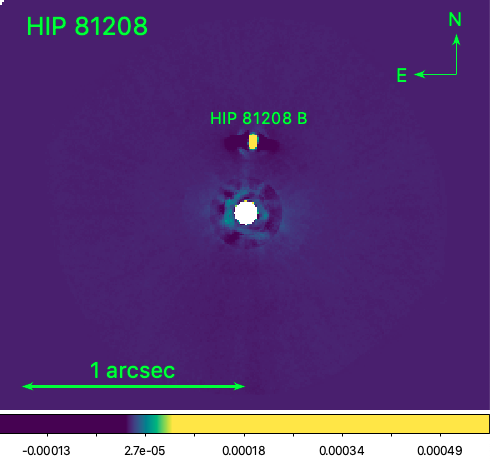} 
  \caption{}
  \label{fig2a}
\end{subfigure}%
\begin{subfigure}{.5\textwidth}
  \centering
  \includegraphics[width=1.\linewidth]{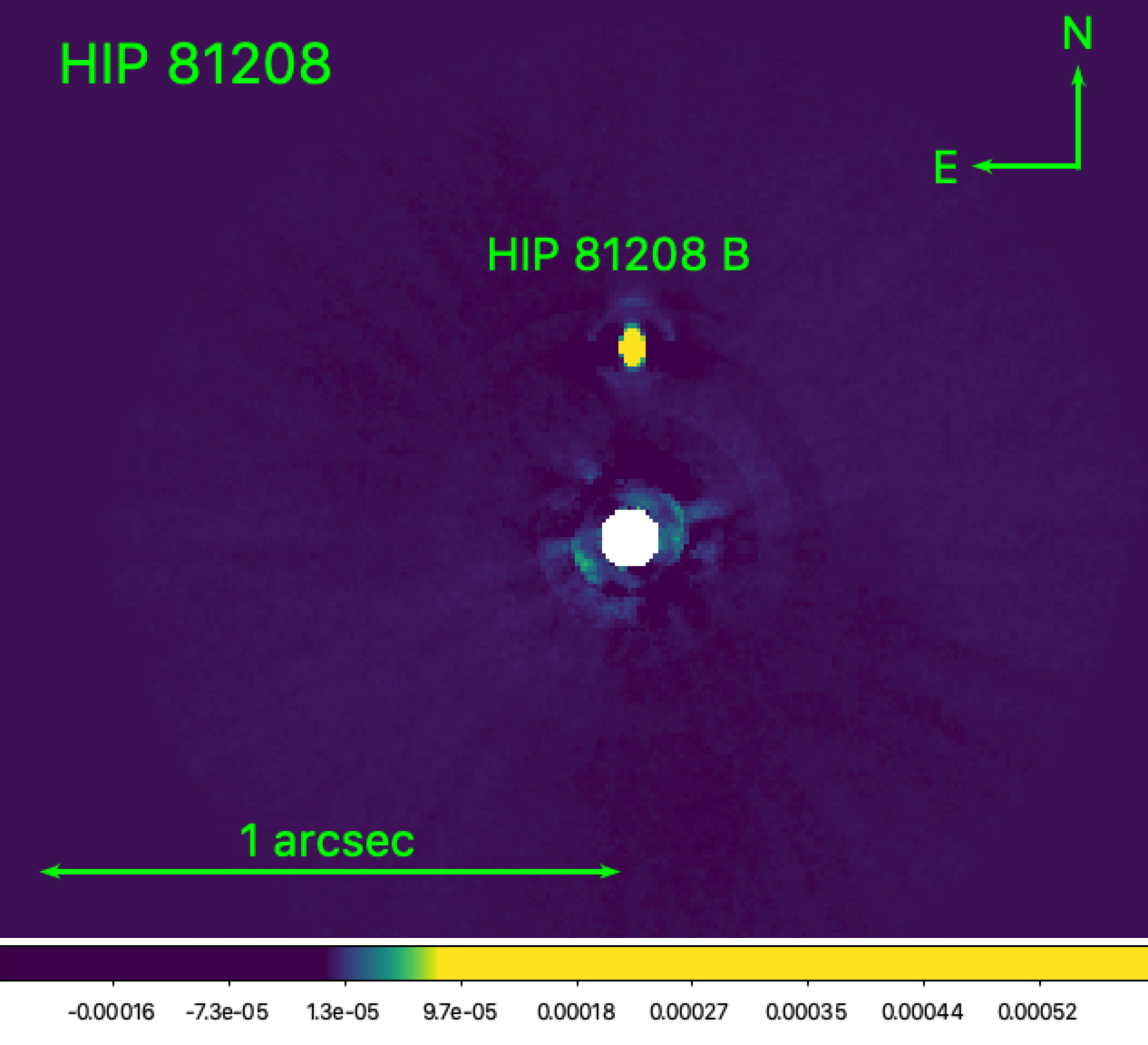}
  \caption{}
  \label{fig2b}
\end{subfigure}
 \caption{SPHERE IFS $YJH$ band image of HIP 81208 from (a) 2019 epoch and (b) 2022 epoch, reduced using a KLIP-based pipeline. The image here is the median taken over the entire wavelength band and shows a bright companion, HIP 81208 B, detected at $0.325\pm0.001$\arcsec~North of the primary star (averaged over the two epochs). The pixel scale is $7.46\pm0.02$ mas/pixel in both epochs.}
\end{figure*}

\begin{figure*}
\centering
\begin{subfigure}{.5\textwidth}
  \centering
  \includegraphics[width=0.96\linewidth]{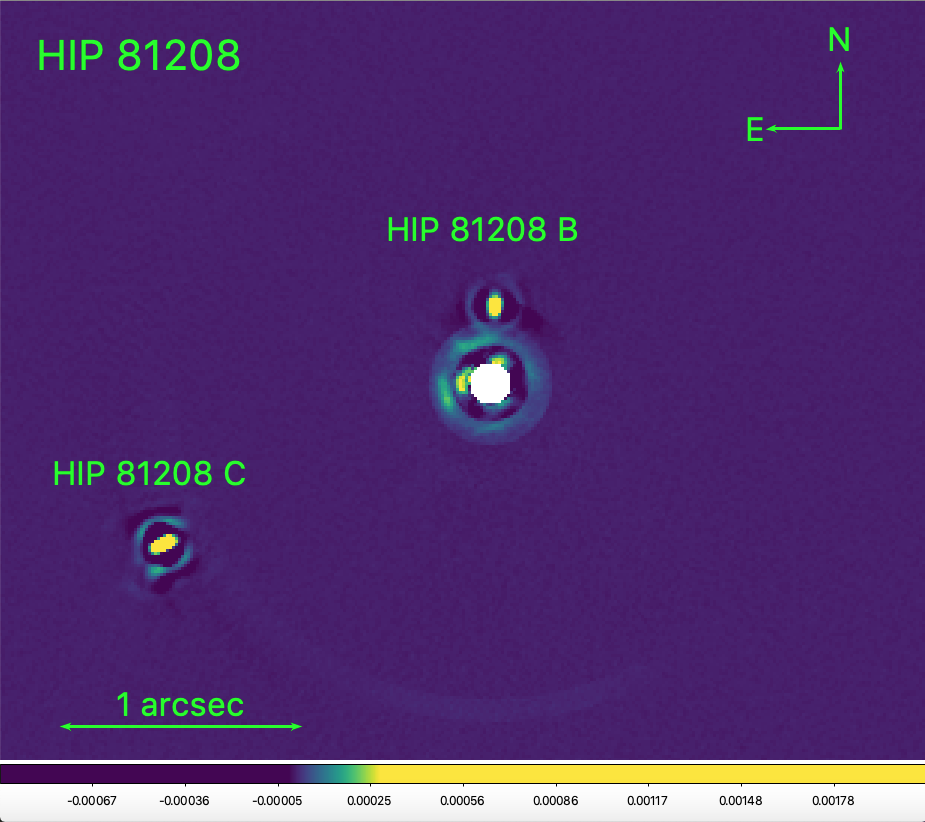} 
  \caption{}
  \label{fig3a}
\end{subfigure}%
\begin{subfigure}{.5\textwidth}
  \centering
  \includegraphics[width=1.\linewidth]{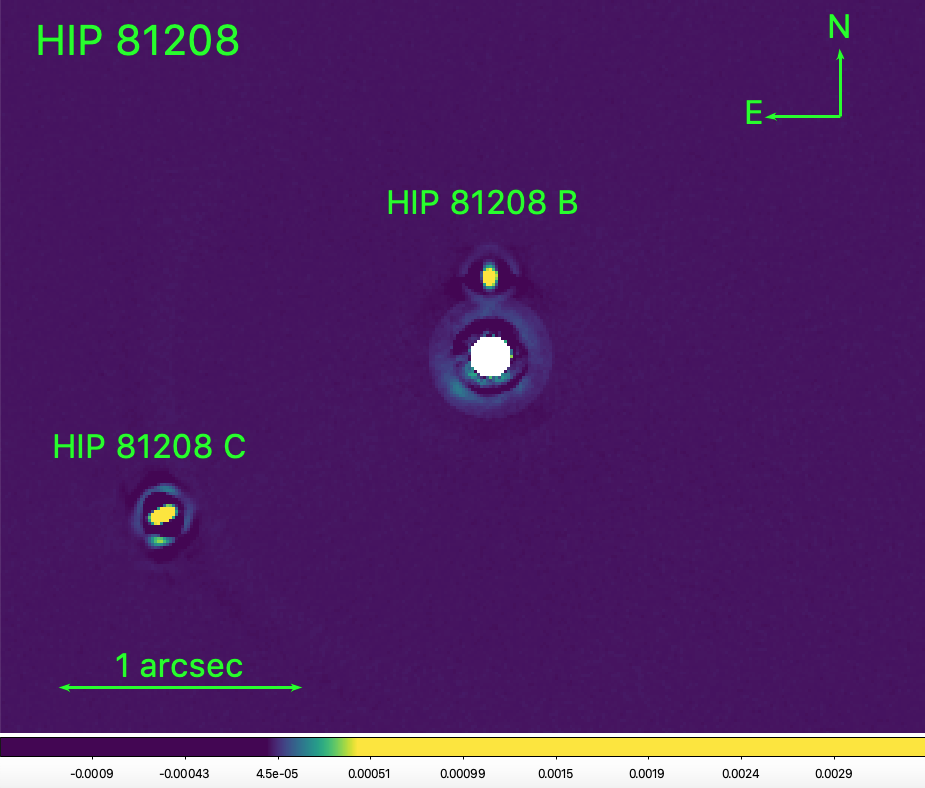}
  \caption{}
  \label{fig3b}
\end{subfigure}
 \caption{SPHERE IRDIS $K$ band image of HIP 81208 from the (a) 2019 and (b) 2022 epoch observations, reduced using SpeCal in TLOCI mode. Each image here is the median taken over $K1$ and $K2$ band images of the respective epoch. Images show the companion HIP 81208 B, as well as a second companion HIP 81208 C, detected at $0.325\pm0.001\arcsec$~North and $1.492\pm0.001\arcsec$~ South East of the primary star respectively (averaged over the two epochs). The pixel scale was $12.258\pm0.004$ mas/pixel for $K1$ band and $12.253\pm0.003$ mas/pixel for $K2$ band.}
\end{figure*}

Post-processing of the IFS image using ADI reveals a candidate companion, HIP 81208B, at a projected separation of $0.325\pm0.001$\arcsec~($48.33\pm0.48$ au; averaged over the two epochs) North of the primary star. The candidate is almost equally bright in both the epochs ($\Delta J = 7.6\pm0.1$ mag), as can be seen in Fig. \ref{fig2a} and \ref{fig2b} which display the median of the reduced images over the entire IFS $YJH$-band in the 2019 and 2022 epochs respectively. The candidate is also detected in the reduced data from IRDIS $K1$ ($\Delta K1 = 6.9\pm0.1$ mag) and $K2$ ($\Delta K2=6.6\pm0.1$ mag) bands in both epochs (refer Fig. \ref{fig3a} and \ref{fig3b}). In addition, a second candidate companion, HIP 81208C, is also revealed in the IRDIS images at a projected separation of $1.492\pm0.001\arcsec~ (221.86\pm2.09$ au, averaged over the two epochs)~from the central star.
This candidate is not within the field of view (FoV) of the IFS ($1.73\arcsec\times1.73\arcsec$).

\begin{figure}
    \centering
    \includegraphics[width=1.0\hsize]{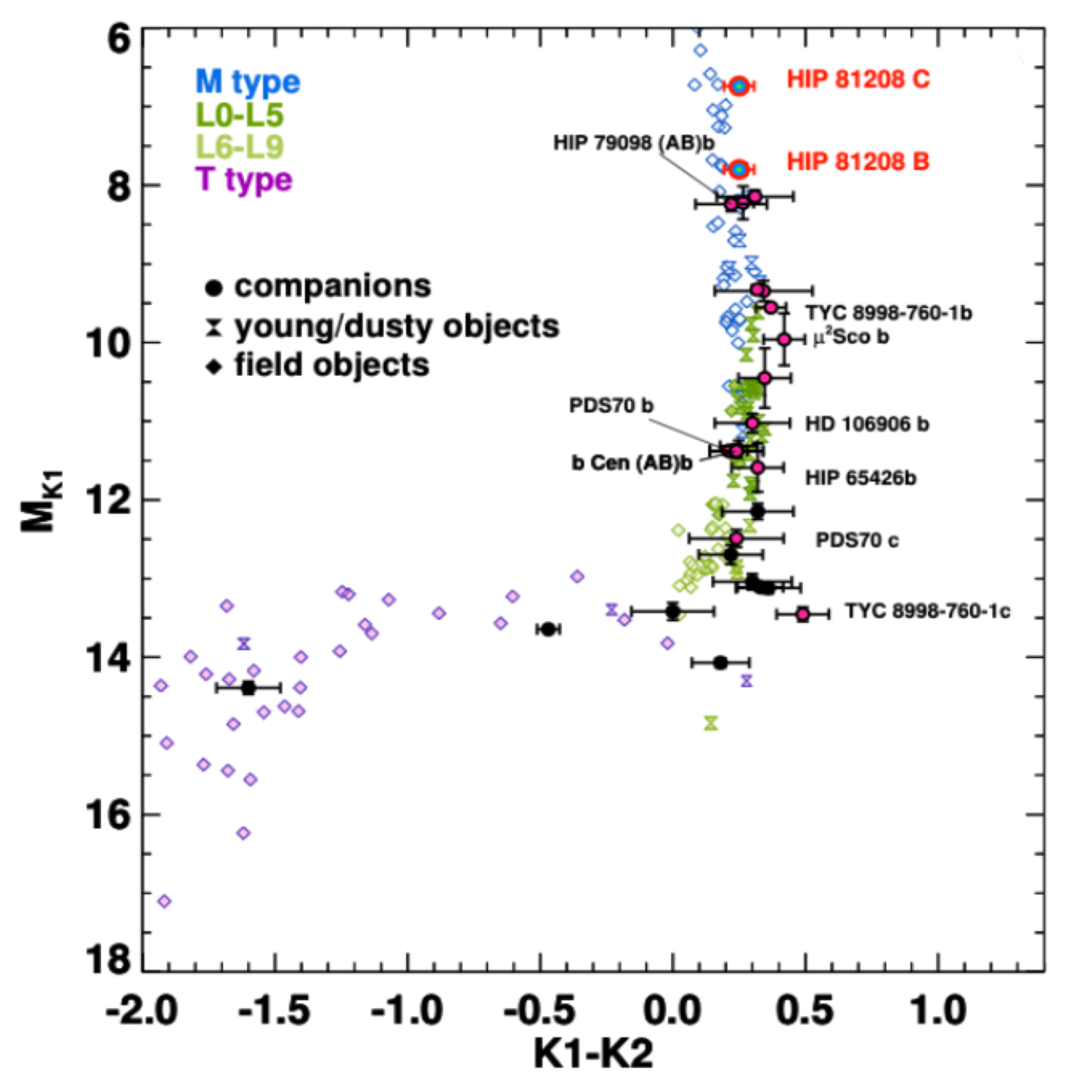}
    \caption{$K1-K2, K1$ colour magnitude diagram, with the candidates HIP 81208B and HIP 81208C shown along with other known sub-stellar companions, young/dusty objects and field stars.}
    \label{fig:cmd}
\end{figure}

Table \ref{tab2} lists the obtained photometry and astrometry for HIP 81208 B and C from the observations in both the epochs. To put both the candidates in the same astrometric reference frame, the astrometry was calculated based only on the IRDIS data. The astrometry in the individual IRDIS bands is obtained by inserting a negative point spread function (PSF) in the approximate position of the planet, and minimising the root mean square (rms) of the residuals in an area of $\sim1\lambda/D$ around this position in the reduced data, by appropriate changes in both the exact position where the PSF was inserted and the multiplicative factor needed to reproduce its intensity. The final astrometry values listed for the companion are the weighted mean of its values from $K1$ and $K2$ bands.
The contrast magnitudes for the companions were obtained using a similar procedure, but keeping the position of the companion fixed at the value obtained from the astrometric analysis. For IFS photometry, all the individual 39 channels were considered separately (see Section \ref{sec:spec} for more details) and the contrast magnitudes in $Y$, $J$, $H$ bands were obtained by taking the weighted mean of the respective values over the spectral channels with wavelength between 1—1.1 $\mu m$, 1.16—1.33 $\mu m$ and 1.5—1.64 $\mu m$ respectively. Fig. \ref{fig:cmd} shows the position of the two candidate companions in the $K1-K2,K1$ colour magnitude diagram, along with other known sub-stellar companions, young/dusty objects and some field stars.

\begin{table*}[!htbp]
\centering
\scalebox{1.0}{
\[
\begin{array}{l|ll|ll}
\hline \hline
\multicolumn{1}{c}{} & \multicolumn{2}{c}{\text{First epoch}} & \multicolumn{2}{c}{\text{Second epoch}} \\
\hline
\multicolumn{1}{c}{} & \multicolumn{1}{c}{\text{HIP 81208B}} & \multicolumn{1}{c}{\text{HIP 81208C}} & \multicolumn{1}{c}{\text{HIP 81208B}} & \multicolumn{1}{c}{\text{HIP 81208C}} \\
\hline
\text{Separation [mas]} & 320.9\pm1.0 & 1493.4\pm1.2 & 328.7\pm1.0 & 1490.0\pm1.8 \\
\text{Position angle [deg]} & 356.55\pm1.72 & 116.26\pm0.08 & 0.43\pm0.13 & 115.96\pm0.08 \\ 
\text{$\Delta Y$ [mag]} & 8.13\pm0.16 & \text{\textemdash} & 8.10\pm0.16 & \text{\textemdash} \\ 
\text{$\Delta J$ [mag]} & 7.60\pm0.12 & \text{\textemdash} & 7.57\pm0.12 & \text{\textemdash} \\ 
\text{$\Delta H$ [mag]} & 7.19\pm0.12 & \text{\textemdash} & 7.19\pm0.13 & \text{\textemdash} \\ 
\text{$\Delta K1$ [mag]} & 6.88\pm0.05 & 5.83\pm0.05 & 6.85\pm0.12 & 5.77\pm0.12 \\ 
\text{$\Delta K2$ [mag]} & 6.64\pm0.07 & 5.59\pm0.07 & 6.59\pm0.12 & 5.51\pm0.12 \\ 
\text{$M_J$ [mag]} & 8.44\pm0.14  & \text{\textemdash} & 8.41\pm0.17 & \text{\textemdash} \\
\text{$M_H$ [mag]} & 8.07\pm0.15 & \text{\textemdash} & 8.07\pm0.15 & \text{\textemdash} \\
\text{$M_{K1}$ [mag]} & 7.76\pm0.09 & 6.71\pm0.09 & 7.73\pm0.14 & 6.65\pm0.14 \\
\text{$M_{K2}$ [mag]} & 7.52\pm0.10 & 6.47\pm0.10 & 7.47\pm0.14 & 6.38\pm0.14 \\
\hline        
\end{array}
\]
}
\caption{Astrometry of HIP 81208 B and C as obtained from IFS and IRDIS data in the first (2019) and second (2022) epochs, along with the obtained contrast magnitudes in the $Y$, $J$, $H$, $K1$ and $K2$ photometric bands. The conversion of contrasts to absolute magnitudes listed in the table for $J$, $H$, $K1$, $K2$ bands have been mediated by photometric data for the primary listed in Table \ref{tab1}.}
\label{tab2}
\end{table*}

\subsection{Confirming the physically bound nature of the candidates}
\begin{figure}
\centering
\includegraphics[width=1.05\hsize]{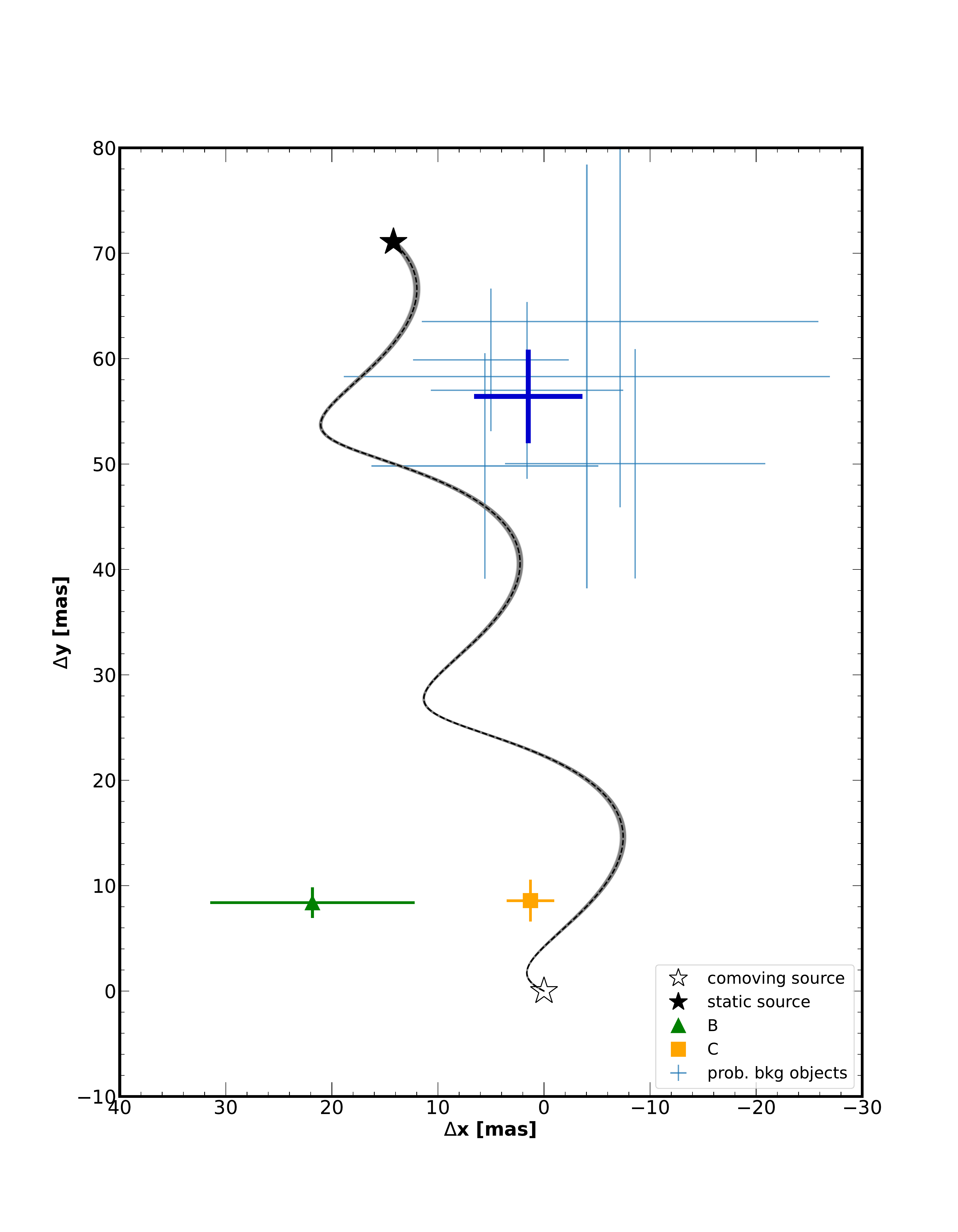} 
\caption{Relative astrometric shifts between the two epochs along $\alpha^*$ (x axis) and $\delta$ (y axis) for all the sources detected in the IRDIS FoV. The black dotted curve shows the expected motion of a pure background source with null proper motion when compared to a source co-moving with the target (the grey shaded region encompasses the uncertainty associated with this motion). The mean astrometric shift of the probable background objects (blue crosses) detected in the data are shown as a thick blue cross, and that of HIP 81208 B and C are shown as a solid green triangle and a solid yellow square respectively.}
\label{fig7}
\end{figure}

To confirm if the directly imaged candidates are physical companions to the primary, we study their proper motion \citep[see, e.g.,][]{janson2021a}. A physical companion to the star will have a similar proper motion to it, but different from those of other background sources in the FoV. Aside from the primary star and the two companion candidates HIP 81208 B and C, we identify 6 other sources detected within the IRDIS ($K1$, $K2$) FoV, that are common to the data in both epochs. 5 additional sources are also seen, but only in one epoch, and are therefore neglected in the following analysis.
We study the astrometric motion of all the above objects relative to the star, between the two epochs (given in Table \ref{table:cc_info} in the Appendix, along with the obtained photometry), as shown in Fig. \ref{fig7}. Given the magnitude of the two candidates B and C, if they are not physically bound to the target star, they would need to be background stars and would in that case, possess a low proper motion tending towards zero, i.e. a null proper motion. Then their astrometric positions would shift between the two epochs as a reflection of the proper motion of the star. This astrometric motion can be predicted and is shown as the black dotted curve in the figure, starting at the position expected for a source co-moving with HIP 81208 (unfilled star) and ending at the position expected from the astrometric shift of a pure background source with null proper motion in the second epoch (black filled star).
Also plotted in the figure are the relative astrometric shift of the 6 additional sources detected in the IRDIS data in both the epochs (blue crosses, labelled as \textit{``prob. bkg objects"}) along with the associated error bars. These shifts are large with respect to the predicted motion of a comoving source; the offsets from the position expected for a completely static source are due to their own proper motion, which could manifest a reflection of the galactic rotation curve. It can be seen that the motion of the two companion candidates B (green cross) and C (yellow cross) are clearly different from those of these 6 sources, and much closer to that of a comoving source. The mean astrometric shift of the 6 additional sources in the IRDIS data is $-12.7 \pm 1.5$ mas (rms: 5.1 mas) along $\alpha$cos($\delta$) and $-14.7 \pm 1.3$ mas (rms: 4.4 mas) along $\delta$, plotted as the thicker blue cross; on the other hand the astrometric shift along $\alpha$cos($\delta$), $\delta$ of HIP 81208 B and C are ($7.6 \pm 9.6$, $-62.7 \pm 1.4$) mas and ($-12.9 \pm 2.2$, $-62.5 \pm 2.2$) mas respectively. This correspondingly places a confident 10.4 $\sigma$ and 9.8 $\sigma$ distinction between the astrometric motion of B and C from the mean astrometric motion of the cloud of background objects detected in the IRDIS field, strongly indicating that these candidates are physically bound to the primary star.

\begin{figure}
\centering
\includegraphics[width=1.1\hsize]{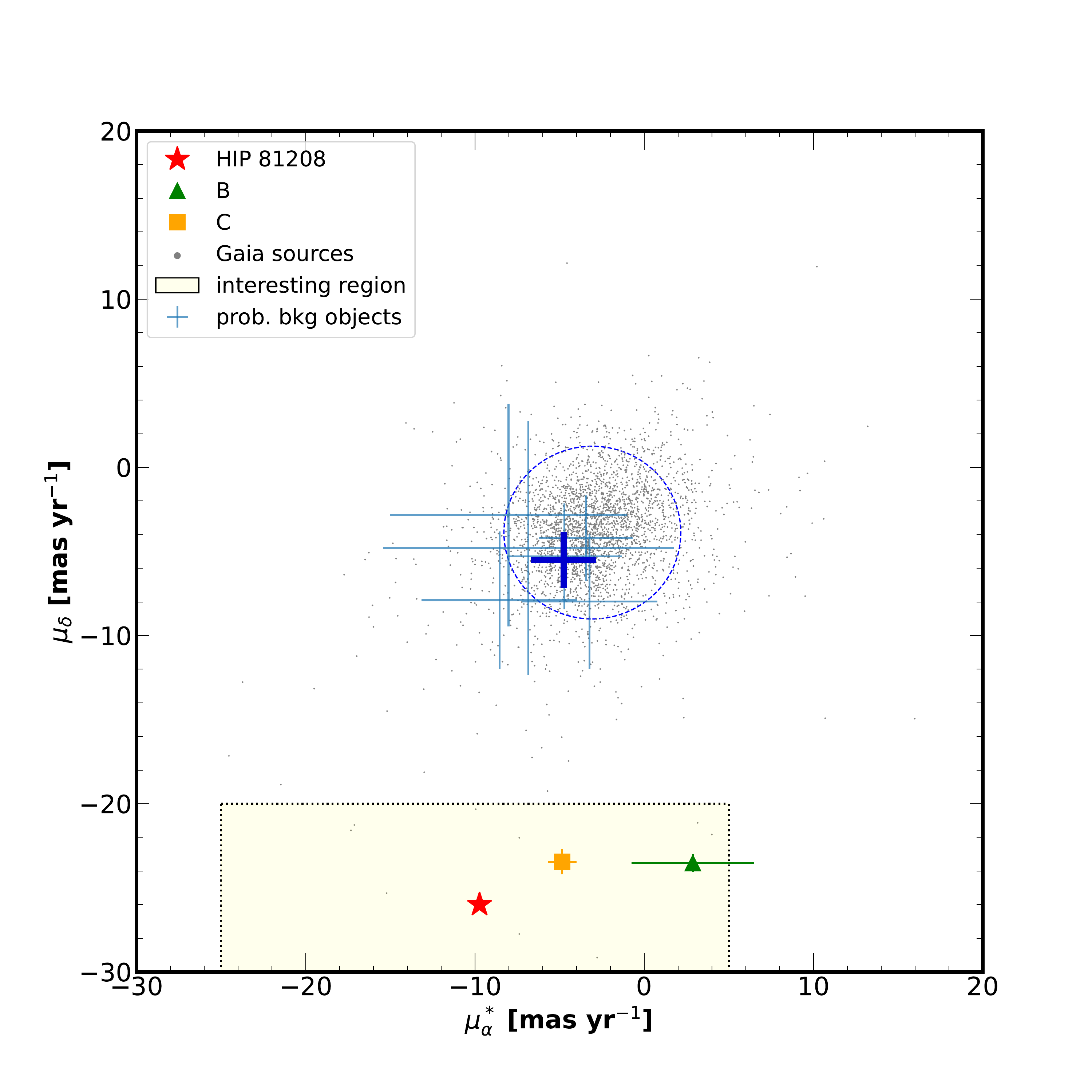} 
\caption{The proper motion in $\alpha$cos($\delta$), $\delta$ for the sample of Gaia EDR3 sources within 5 arcmin from the primary, and for the 6 background sources in the IRDIS data. Also shown are the median proper motion of the 6 probable background objects as well as the proper motion derived for HIP 81208 B and C using the same symbols as in Fig. \ref{fig7}. The interesting region of proper motion defined by Equation \ref{eq2} is over-plotted in the figure as a yellow-shaded region enclosed within black dashed contours.}
\label{fig8}
\end{figure}

To quantify this indication, we compute the probability that the two companion candidates, if drawn from such a sample of background sources could possess an astrometric shift which would mimic that of a physically bound companion of the primary. In order to compute such a false alarm probability (FAP), we queried the Gaia EDR3 catalogue \citep{gaia2020} for field stars near HIP 81208, with a search radius of 5 arcmin. The resulting sample contains 4014 Gaia sources; the 50$^{\text{th}}$ percentile of their proper motion distribution (with the 16$^{\text{th}}$ and 84$^{\text{th}}$ percentiles shown as the lower and upper uncertainty limits) are $<\mu_{\alpha^*}>=-3.12_{-2.71}^{+3.11}$ \masyr and $<\mu_{\delta}>=-3.67_{-2.85}^{+2.57}$ \masyr . These are shown as the grey dots in Fig. \ref{fig8}, labelled as "Gaia sources". In the same figure, we also show the 6 background sources from IRDIS data as blue crosses, with their proper motion computed as the ratio between the corresponding astrometric shifts with respect to the position expected for a static source and the time baseline between the two epochs. The mean proper motion of these 6 sources, along with the associated uncertainty, is $<\mu_{\alpha^*}>=-4.77 \pm 0.58$ \masyr~and $<\mu_{\delta}>=-5.50 \pm 0.50$ \masyr~(shown as the thicker blue cross in the figure), which is well within  $\sim 1\sigma$ of that of the Gaia sample.
Similarly, the derived  $\mu_{\alpha^*}$, $\mu_{\delta}$ of HIP 81208 B and C are ($2.87 \pm 3.62$, $-23.53 \pm 0.54$) \masyr and ($-4.85 \pm 0.84$, $-23.46 \pm 0.75$) \masyr, respectively. To estimate the FAP, we define an "interesting region" of proper motion in the figure, similar to \cite{squicciarini2022a}, within which a background star might be disguised as a physically bound source. This region employs the following boundaries:

\begin{equation}
    \begin{cases}
    -25~\text{\masyr} < \mu_{\alpha^*} < 5~\text{\masyr},\\
    \mu_{\delta} < -20~\text{\masyr}
    \end{cases}
    \label{eq2}
\end{equation}

The number of Gaia stars that fall into this region is 12 out of 4014 stars, resulting in a corresponding fraction of "interesting objects" being $\sim3\times10^{-3}$. Given that there are 8 source detections in the entire IRDIS FoV, this gives an FAP of $\sim$2\%. The probability of having at least two such sources is as low as $2 \times 10^{-4}$. The obtained low value of FAP is sufficient enough to add further robustness to the claim that HIP 81208 B and C are physically bound companions to the primary.

It is also evident from Table \ref{table:cc_info} that both the companions B and C are much brighter than all other sources detected in the IRDIS FoV. An average over the $K1$, $K2$ bands over the two epochs would yield $K=13.51\pm0.06$ mag for B and $K=12.44\pm0.06$ mag for C. An alternative way to determine if HIP 81208 B and C are background objects would then be to compute the probability that background objects as bright as these can be detected at such small separations from the star. Since the two companions are quite bright, we may use the 2MASS catalogue \citep{skrutskie2006}, which is complete for uncrowded regions at this magnitude level, to estimate the surface density of background objects as bright as B and C. Searching within a radius of 2 arcmin from HIP 81208, we found 26 and 15 2MASS sources brighter than B and C respectively. Given the projected separation of the two companions from the star, this corresponds to very low probabilities of $1.9\times10^{-4}$ and $2.3\times10^{-3}$ that B and C, respectively, are background objects by chance projected very close to HIP 81208. 

An underlying possibility that is yet to be investigated is that the two companions, even though co-moving as established above, could be UCL members that are not bound to the target. Considering the derived masses of B and C, we first evaluate the probability that B is a free-floating UCL substellar object that was by chance close to the target. For this, we follow a similar analysis by \cite{squicciarini2022a} and integrate the normalised IMF of the UCL association from 5 $M_J$\footnote{The lower limit 5 $M_J$ corresponds roughly to the sensitivity of BEAST survey, given the performances of SPHERE and the age and distance of Sco-Cen.} to 75 $M_J$ to obtain the fraction of objects in UCL belonging to this mass range. To obtain the expected number of UCL objects between 5—75 $M_J$, we multiply this fraction by the number of UCL sources ($\sim$4021)\footnote{Rescaled, considering the limiting magnitude of \textit{Gaia}, from the actual list of members in \citep{Damiani2019} which was complete only above 15 $M_J$.}. We further divide this value by the area of UCL sub-region as given in \cite{squicciarini2022a} to estimate the corresponding projected density of these objects. Multiplying the projected density with the IRDIS FoV (11\arcsec$\times$11\arcsec) gives us the number of free-floating substellar objects expected in the IRDIS FoV. Using a binomial distribution, we can then get the FAP of having seen atleast one such object across the 47 targets that have been observed atleast twice in the BEAST survey as $\sim5\times10^{-4}$. Similarly, we also computed the probability that C is a stellar mass UCL member that is not bound to HIP 81208, by repeating the above calculation but considering every possible companion mass from 5~$M_J$ to the most massive UCL star $\sim$15 $M_{\odot}$. The corresponding value for FAP is $\sim2.4\times10^{-3}$. This probability can be regarded as an upper limit since this analysis considers all the stars in UCL, but the actual number of interloping stars we should estimate in the IRDIS FoV should not have those stars considered, for which \textit{Gaia} has resolved the true position. So we can safely say that the probability of seeing by chance, due to projection effects, at least one UCL member of any stellar mass $>5~M_J$ across all the stars that have been observed twice in BEAST survey until now is $<0.2-0.3\%$.

\subsection{Characterisation of Candidate Properties}
\label{sec:mass}
Photometric mass estimates were derived for both the candidates with the help of \textsc{madys}, based on the average of their respective contrast measurements over the two epochs (refer Table \ref{tab2}). In particular, the estimate for HIP 81208C is based on IRDIS ($K1$, $K2$) bands while that for HIP 81208B is based on both IRDIS and synthetic IFS magnitudes $J_{IFS}$ at $1.246~\mu$m (band width=0.174~$\mu$m) and $H_{IFS}$ at $1.570~\mu$m (band width=0.132~$\mu$m), which were obtained by collapsing the spectral channels 12-21 (1.159-1.333 $\mu$m) and 30-38 (1.504-1.636 $\mu$m), respectively. The derivation of calibrated apparent magnitudes from contrasts is mediated by suitable photometric data for the primary, taken from the literature: 2MASS  $J$, $H$ and $K_s$ for IFS $J_{IFS}$, $H_{IFS}$ and the doublet ($K1$, $K2$), respectively (refer Table \ref{tab1}). Given the spectral classification of the primary as a B9V star, the errors introduced by these approximations are well within photometric uncertainties.

\textsc{madys} operates by seeking the best match between the vector of input photometry for the object and a selected track/isochrone grid via $\chi^2$ minimisation. Evaluation of random uncertainties associated to apparent photometry, parallax, extinction, and age is naturally taken into account in this process in a Monte Carlo fashion. In order to evaluate the impact of theoretical uncertainties on the results, we repeated the mass estimation of each companion twice, employing a different suite of stellar/substellar evolutionary model each time; namely the Ames-Cond models \citep{allard2001} and the BHAC15 models. For each candidate, the individual estimates from both repetitions were consistent with one another, and were averaged to get the final mass estimate:

\begin{equation}
M_B = 67^{+6}_{-7}~M_J = 0.064^{+0.006}_{-0.007} M_\odot,
\end{equation}

\begin{equation}
M_C = 141^{+10}_{-14}~M_J = 0.135^{+0.010}_{-0.013}~M_\odot
\end{equation}

The corresponding \teff~returned by the method as a result of the above interpolation scheme were also averaged to get <\teff> $=2895^{+45}_{-40}$ K for B and <\teff> $=3165^{+40}_{-60}$ K for C. As indicated from Table 6 of \cite{PM2013}, listing intrinsic colours, adopted \teff~ and bolometric corrections of 5-30 Myr stars, these values roughly correspond to a spectral type of M5 for B and M4 for C. A comparison of the \teff~ estimate for B can be made using its observed spectrum over the IFS and IRDIS bands (see Section \ref{sec:spec}), giving some level of confidence on the derived \teff~and spectral type. However, no such comparison can be made for C since it lacks an observed spectrum in the IFS, so the derived \teff~and spectral type for C here should be subjected to careful interpretation.

\cite{grieves2021} briefly summarises the stellar-substellar mass boundary predictions in the literature. Although the hydrogen-burning mass limit is generally adopted as 80 $M_J$, the boundary predictions range from 73.3--98.5 $M_J$ depending on the choice of models and metallicities \citep[see ][]{dietrich2018}. An even lower mass limit of $\sim$70 $M_J$ was predicted by \cite{dupuy2017} based on the astrometric masses of 31 ultracool binaries with component spectral types M7--T5. Comparing the above derived mass estimates for the companions with the stellar-substellar boundary predictions will place HIP 81208C in the stellar regime and HIP 81208B most likely in the brown dwarf regime. However, further observations and more stringent mass constraints are required to definitely confirm whether HIP 81208B is a brown dwarf or a very low mass star.

\subsection{Analysis of HIP 81208B spectrum}
\label{sec:spec}
\begin{figure}
\centering
\includegraphics[width=1\linewidth]{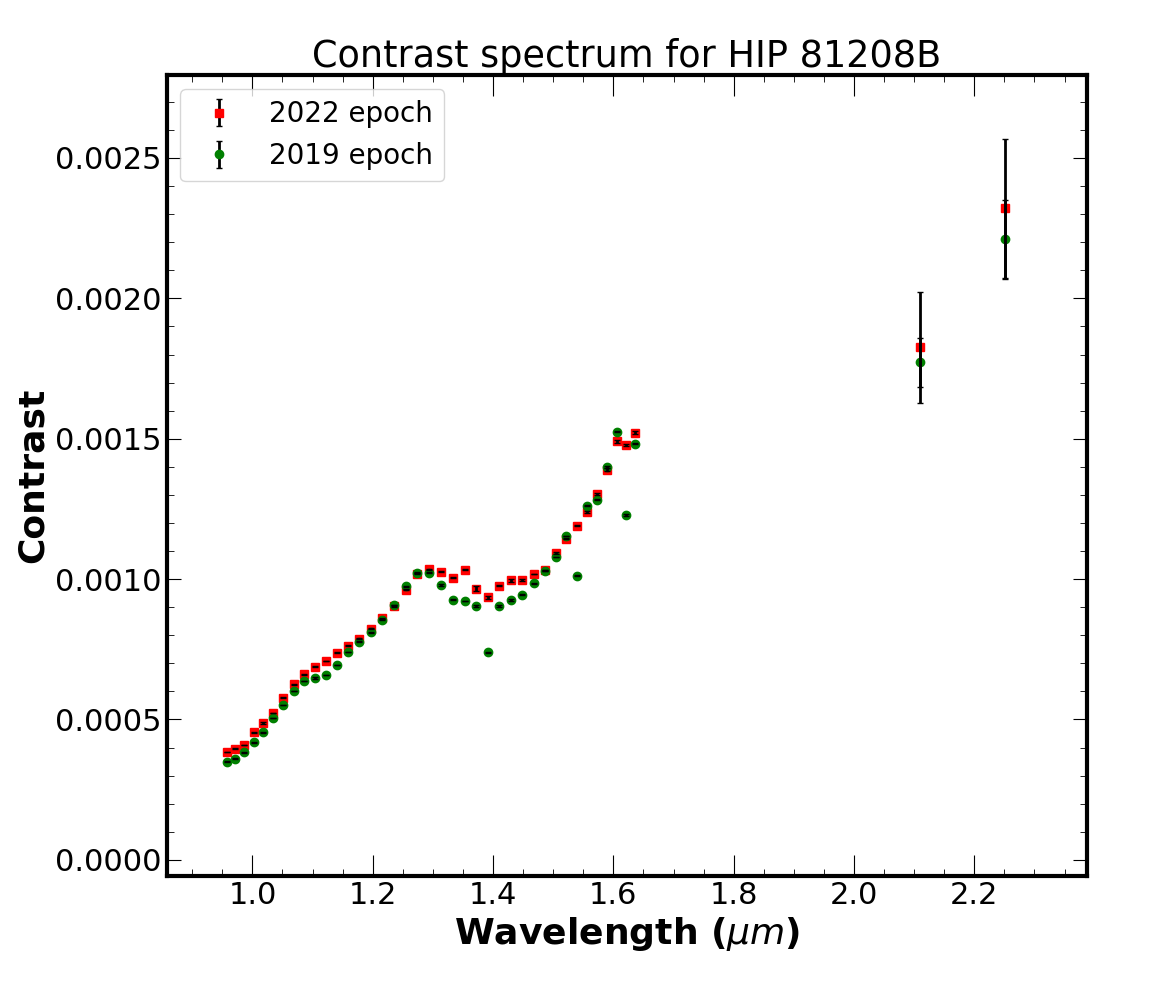} 
\caption{The contrast spectrum for HIP 81208B obtained from reduced IFS data in the $YJH$ band and IRDIS data in $K1$, $K2$ bands, along with the associated uncertainties. The red solid squares represent the contrast obtained from the 2022 observations and the green solid circles represent the contrast obtained from 2019 observations.}
\label{fig4}
\end{figure}

To determine contrast values for HIP 81208B from the KLIP-reduced IFS data, a negative PSF was injected at the position of the candidate.  Monochromatic reductions were then performed based on principal component analysis (PCA) with modes 2--6 in such a way that the standard deviation is minimised within a region of $\sim1\lambda/D$ centred on the candidate's position. The values obtained for each wavelength with the different PCA modes were averaged to get the final contrast for that wavelength. The uncertainty in the contrast for each wavelength was calculated as the rms scatter of the values obtained by repeating the above same procedure, with the negative PSF inserted, not at the companion position, but at five different positions in the image at the same separation from the star as the companion B, and separated by 60, 120, 240 and 300 degrees in position angle. The minimisation of standard deviation only concerns the intensity of the negative PSF that is inserted and hence can be performed even without an actual companion present at the location of insertion. The intensity value derived this way can be either positive or negative, depending on the local noise realisation. The contrast values and uncertainties thus obtained for the candidate from the IFS data in both the epochs are plotted along with the obtained contrasts using SpeCal from the IRDIS $K1$ and $K2$ bands, against the respective wavelengths in Fig. \ref{fig4}.

\begin{figure}
\centering
\includegraphics[width=1\linewidth]{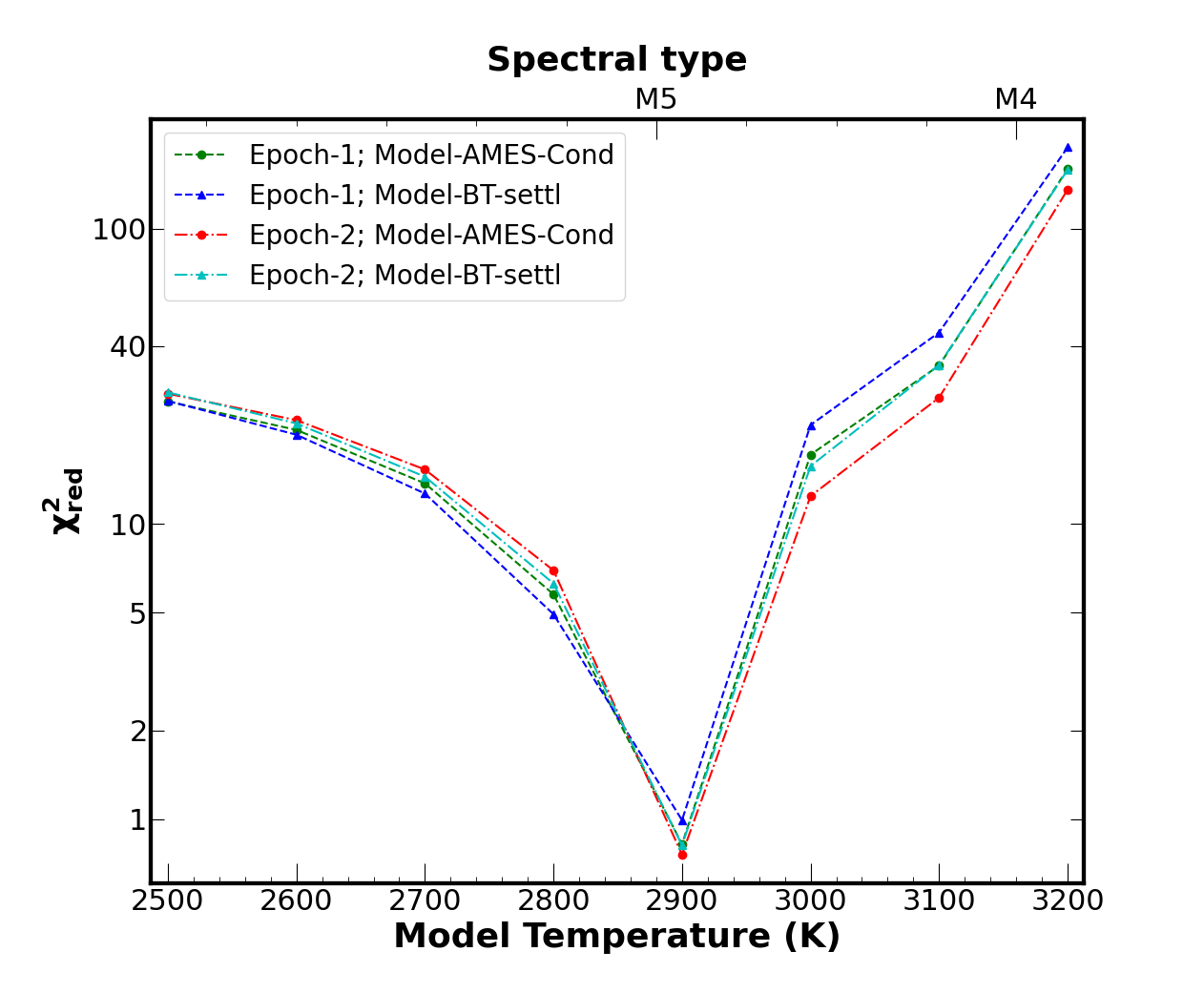} 
\caption{$\chi_{red}^2$ computed between the observed spectrum and the AMES-Cond and BT-Settl models for different effective temperatures, for both 2019 and 2022 epochs. The log(\textsl{g})  adopted for both models is 4.5. For both epochs, the models reach a $\chi_{red}^2\sim1$ at $T_{eff}$=2900 K. The closest spectral type is M5.}
\label{fig5}
\end{figure}

\begin{figure}
\centering
\includegraphics[width=1\linewidth]{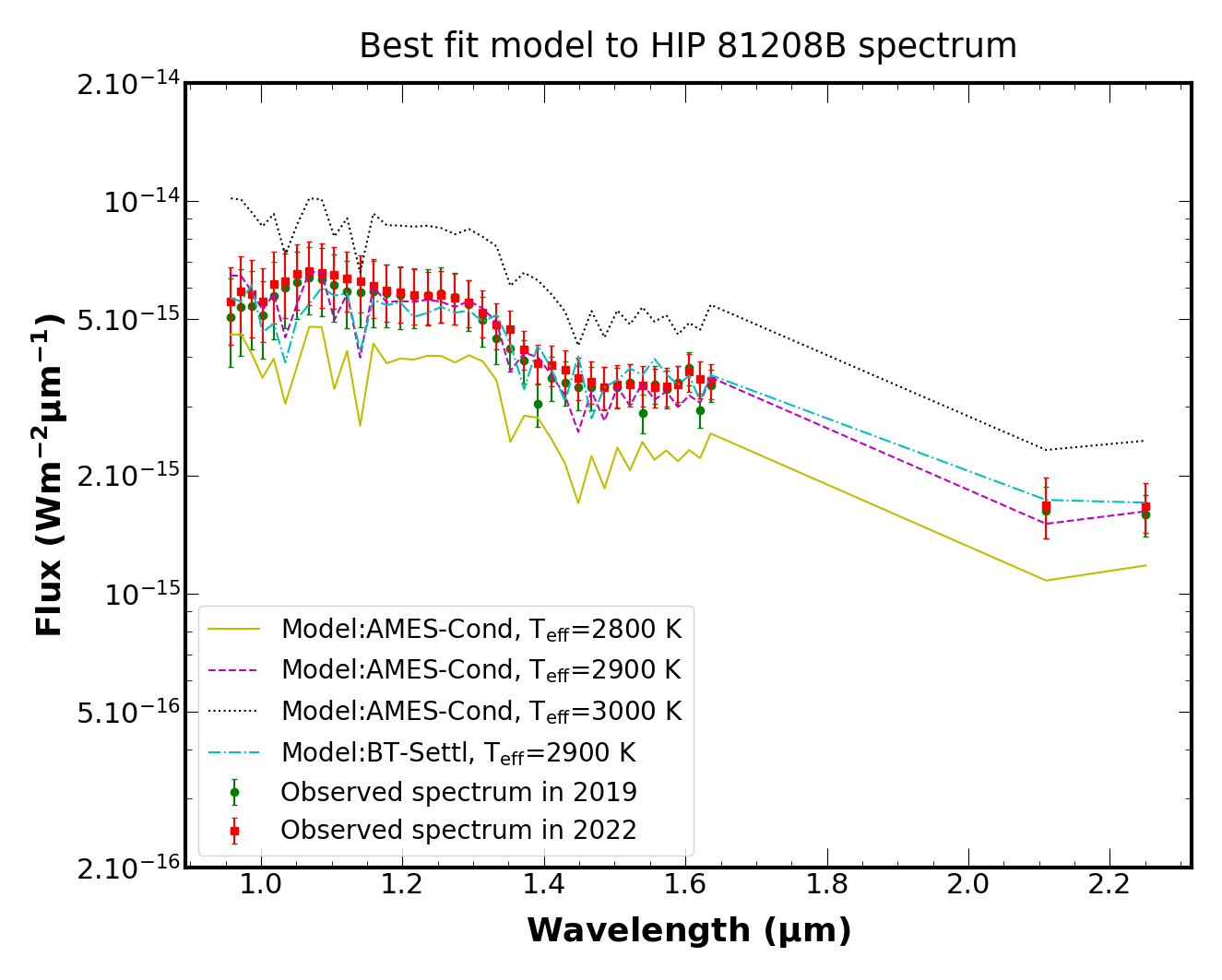} 
\caption{Observed spectrum for HIP 81208B in 2019 (green solid circles) and 2022 (red solid squares) epochs, along with the associated uncertainties in flux. Also plotted are AMES-Cond 2000 log(\textsl{g})=4.5 theoretical model for \teff=2800, 2900 and 3000 K as well as BT-Settl log(\textsl{g})=4.5 theoretical model for \teff=2900 K. The best fit models to the observed spectrum in both epochs correspond to a \teff=2900 K.}
\label{fig6}
\end{figure}

To calculate the flux of the companion B corresponding to the contrast values at each wavelength, we interpolated the best-fit theoretical spectrum for the star from  Section \ref{sec:teff} (see Fig. \ref{fig1}), scaled to the observed flux levels, at the respective wavelengths. The uncertainties in flux are obtained by summing quadratically the uncertainties in contrasts and the photon noise at the location of B estimated within an area of $\sim1\lambda/D$ around it. Further, to determine the spectral type of the candidate, we fit theoretical stellar models AMES-Cond 2000 \citep{allard2001} and BT-Settl \citep{allard2012} to the obtained flux spectrum of the candidate. To determine the right choice for surface gravity, we interpolated the AMES-Cond isochrone corresponding to the estimated age of the target in this work (0.02 Gyr) for the estimated mass of HIP 81208 B in Section \ref{sec:mass} (0.064 M$_{\odot}$); the corresponding log(\textsl{g}) given by the isochrone is 4.42. The closest available choice of surface gravity among the models is log(\textsl{g})=4.5. As an additional check, we also interpolated the AMES-Cond evolutionary track corresponding to a mass of 0.06 M$_{\odot}$; at a log(\textsl{g})=4.5 the age pointed by the evolutionary track is 0.026 Gyr, which is compatible with our age estimate for the target. For these reasons, we chose a log(\textsl{g}) value of 4.5 for the theoretical models in this analysis. The resulting theoretical flux at the surface of the star was scaled to the observed flux level by a factor of $R^2/D^2$, where $R$ is the radius of the candidate for the specific temperature of the model as determined from the corresponding model isochrones and evolutionary tracks corresponding to an age of 0.02 Gyr, and $D$ is the distance to HIP 81208 from Earth. Fig. \ref{fig5} shows the reduced $\chi^2$ ($\chi_{red}^2$) values between the observed spectrum and the two theoretical models at specific effective temperatures. For both 2019 and 2022 epoch observations, $\chi_{red}^2$ value for both models approaches the ideal value 1 at the effective temperature of 2900 K, the closest spectral type to which is M5 ($T_{eff}$=2880 K) \citep[see Table 6 of][]{PM2013}. Fig. \ref{fig6} shows the observed spectrum for HIP 81208B in both the epochs, alongside theoretical models corresponding to the least $\chi_{red}^2$ (2900 K), as well as for \teff=(2800 K, 3000 K). The obtained \teff~ here for B from its spectrum, as well as the corresponding spectral type, is consistent with the results from \textsc{madys} in Section \ref{sec:mass}, rendering these values reliable. We thus conclude that HIP 81208B is likely of spectral type M5.

\subsection{Constraining the orbital parameters}
\label{sec:orbit}
\begin{figure*}
\centering
\begin{subfigure}{.5\textwidth}
  \centering
  \includegraphics[width=0.96\linewidth]{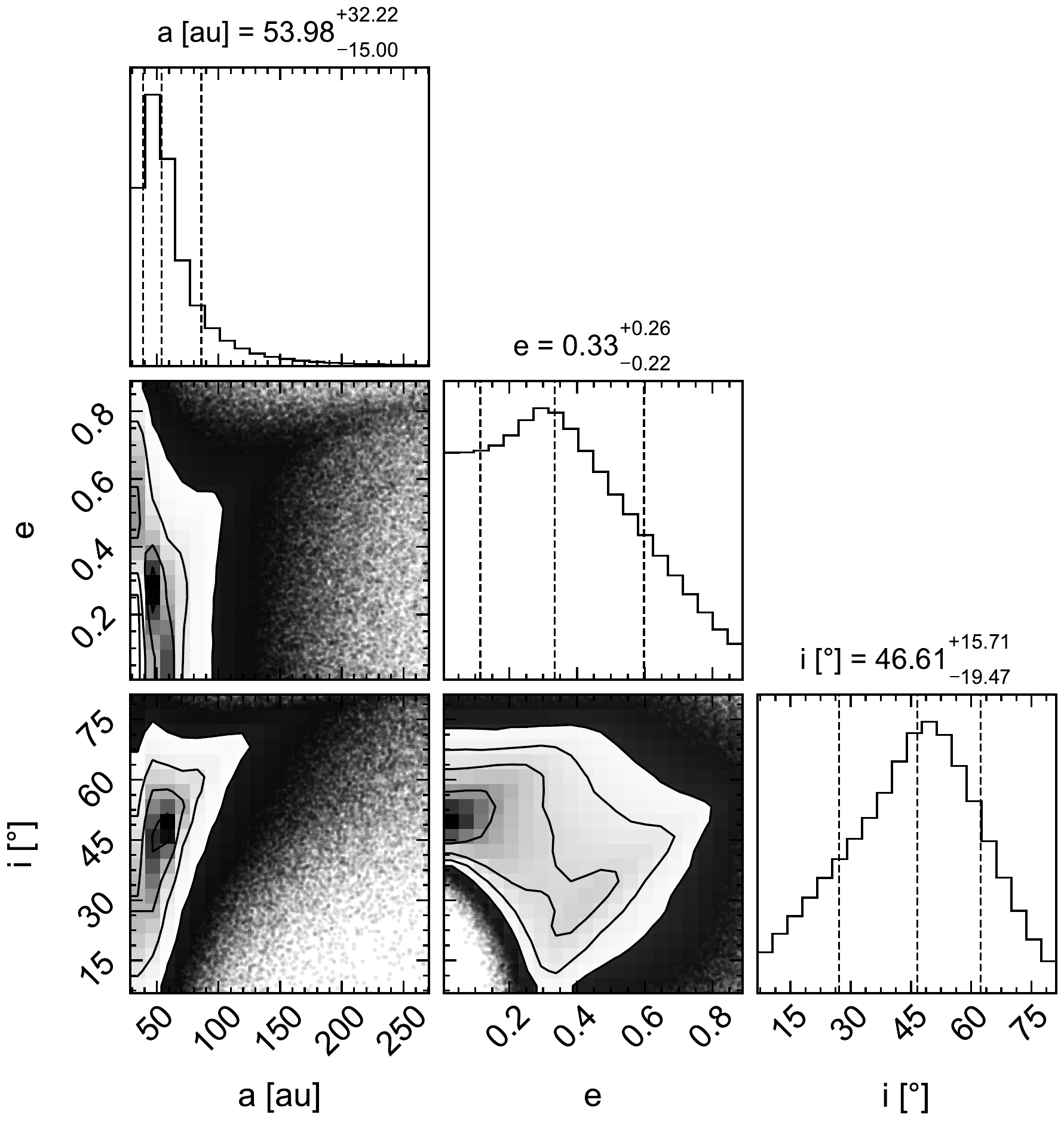} 
  \caption{HIP 81208 B}
  \label{fig9a}
\end{subfigure}%
\begin{subfigure}{.5\textwidth}
  \centering
  \includegraphics[width=0.95\linewidth]{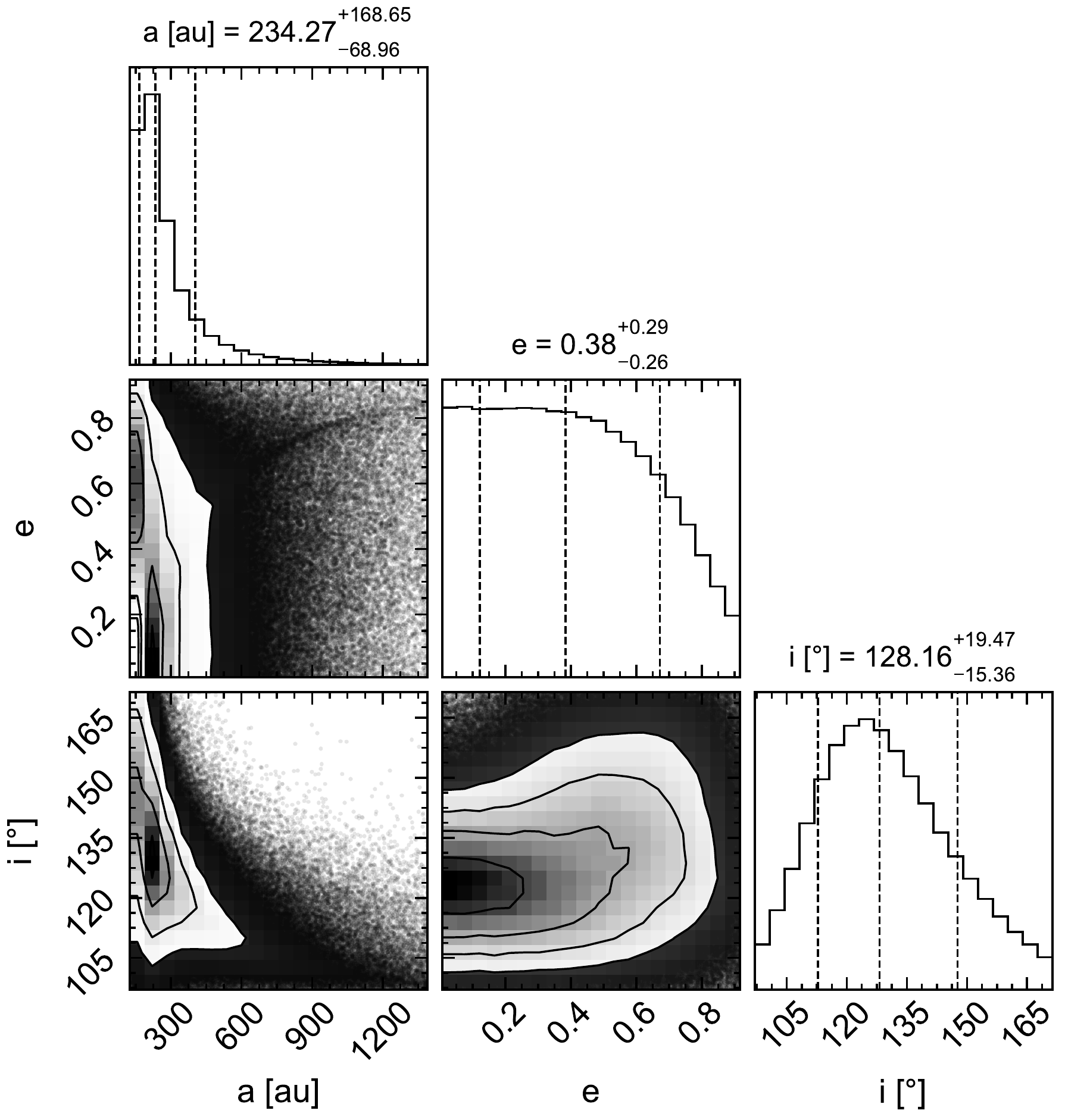}
  \caption{HIP 81208 C}
  \label{fig9b}
\end{subfigure}
 \caption{Corner plots for the accepted orbits from OFTI for the companion B (a) and C (b). The median values along with upper (84\%) and lower (16\%) quantile intervals ($\sim\pm1\sigma$) are shown as dashed vertical lines on the posterior distributions of the respective orbital parameters. The 2-D contour plots show the 1$\sigma$, 2$\sigma$ and 3$\sigma$ confidence limits to these values.}
 \label{fig9}
\end{figure*}

Given the astrometric information of the two companions obtained in two different epochs, it is possible to derive constraints on their orbital parameters. We do this using the python package \textit{orbitize!} \citep{blunt2020}, which is designed to fit the orbits of directly imaged planets. \textit{Orbitize!} offers two choices of algorithm for fitting orbits: Orbits for the Impatient \citep[OFTI, see][]{blunt2017} and Markov Chain Monte Carlo \citep[MCMC, see][]{ford2004, vousden2016}. The disadvantage with using MCMC for orbital fitting for long-period orbits is that for cases with less constraints from observations on the orbital parameters, MCMC take a very long time to converge. Since the relative displacement between the two epochs for B and C were only 7.8 mas and 3.4 mas respectively, we had very low orbital coverage for the companions from our observations, and thus, MCMC would not be an ideal choice. So our choice of algorithm was OFTI, which is a Bayesian Monte Carlo rejection-sampling method that is ideal for cases where the observations cover only a small fraction of long-period orbits. OFTI takes the separation and position angle of the companion in the different observation epochs, the parallax and the total mass of the system (star and companion), along with uncertainties, as input and uses built-in\footnote{See \url{https://orbitize.readthedocs.io/en/latest/tutorials/Modifying_Priors.html} for the complete list of default priors in \textit{Orbitize!} and the different prior choices.} prior probability distribution functions (PDFs) to compute posterior PDFs of orbital parameters. This computation is based on the orbits the algorithm accepts from the generated ones, using the technique of rejection-sampling. We ran \textit{orbitize!} until a total of $10^6$ orbits were accepted by the algorithm for both B and C. The corner plots in Fig. \ref{fig9a} and \ref{fig9b} show the resulting posterior distribution for semi-major axis (\textit{a}), inclination (\textit{i}) and eccentricity (\textit{e}) of these orbits (the plots are truncated to include only 98\% of the generated orbits for better visibility). The 2-D contour plots in the figures illustrate 1$\sigma$, 2$\sigma$ and 3$\sigma$ confidence levels on these values. The median orbital parameters for B and C as derived from these posteriors, along with the upper (84\%) and lower (16\%) quantile intervals ($\sim\pm1\sigma$), are also shown in the figures, as well as listed in Table \ref{tab3}. The orbital periods implied for B and C from their semi-major axis obtained above and their derived masses in Section \ref{sec:mass} are also listed in the table.

\begin{figure}[h]
    \centering
    \includegraphics[width=\linewidth]{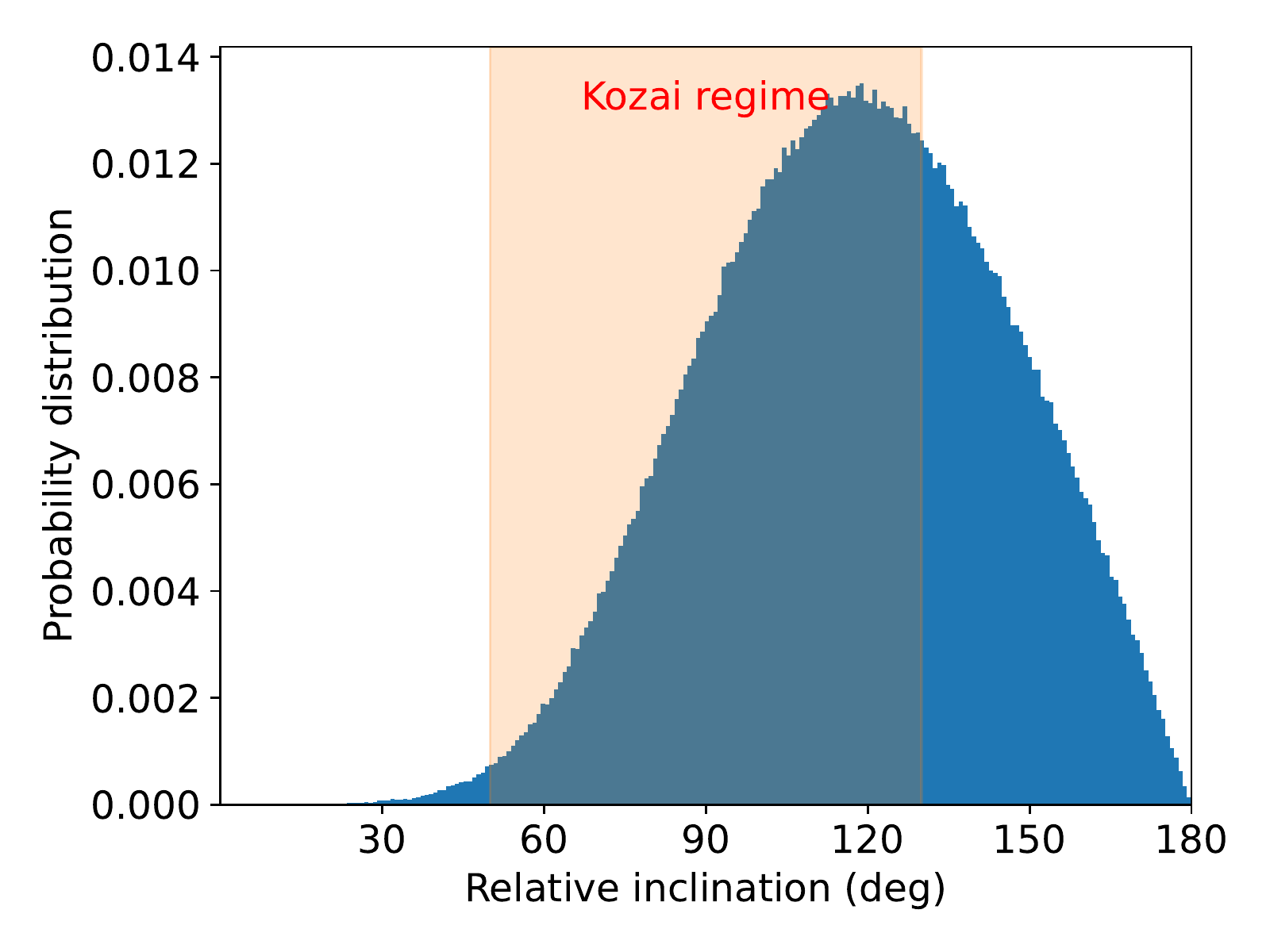}
    \caption{Distribution of the relative inclinations between B and C's orbits from the sample of solutions obtained by the OFTI algorithm. The light red area corresponds to systems in the ZLK configuration, or Kozai regime.}
    \label{fig:inc}
\end{figure}

From the corner plots, it is very unlikely that B has a very high eccentricity orbit (>0.7); higher inclinations (>75$^{\circ}$) than the predicted quantile limit would favour a much wider orbit, both of which seem of very low probability as per the posteriors. Similarly, for C as well, the probability falls rapidly for higher eccentricities and higher values of orbital distance seem very unlikely. It is noteworthy from these results that the orbital inclination of C is very different from that of B, with the relative inclination between the two companions $\sim80^{\circ}$ (refer Fig.~\ref{fig:inc} for the distribution of relative inclinations between the orbits of B and C obtained from the algorithm). Thus, the two orbital planes look roughly orthogonal, making B and C appear to orbit in an opposite sense with respect to each other.

\begin{figure*}
\centering
\includegraphics[width=1\linewidth]{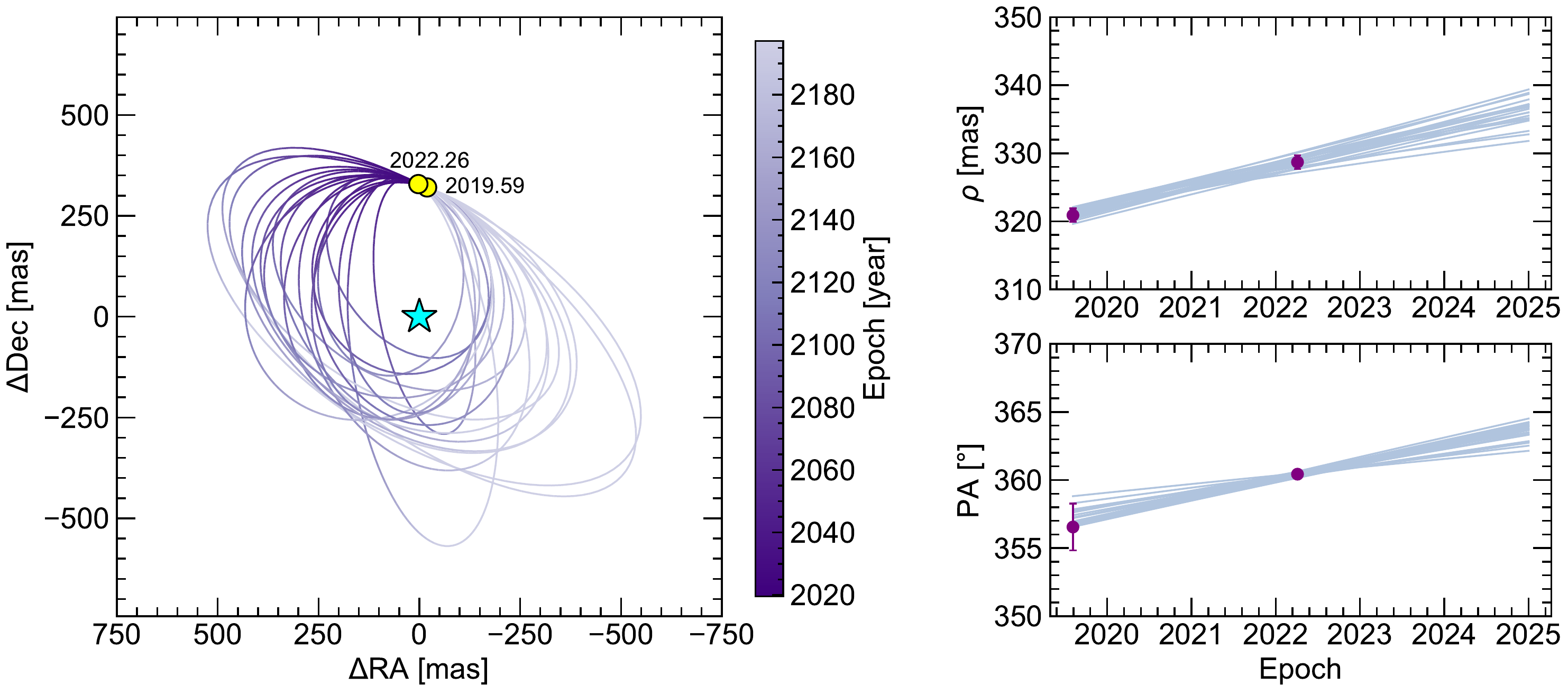} 
\caption{On the left: The motion of HIP 81208 B in RA and DEC over an entire orbital period for 20 random orbits drawn from the posterior distributions from the OFTI run. The adjacent color bar shows the epochs along the orbital positions. The yellow solid circles represent the predicted position of the companion in these orbits in the two observation epochs. The blue star represents the central primary star HIP 81208.  On the right: The projected separation $\rho$ and position angle PA (deg) of B over time, as  predicted from these 20 orbits. The two purple dots on both figures represent the measured values of the respective parameters at the two observation epochs.}
\label{fig10}
\end{figure*}

\begin{figure*}
\centering
\includegraphics[width=1\linewidth]{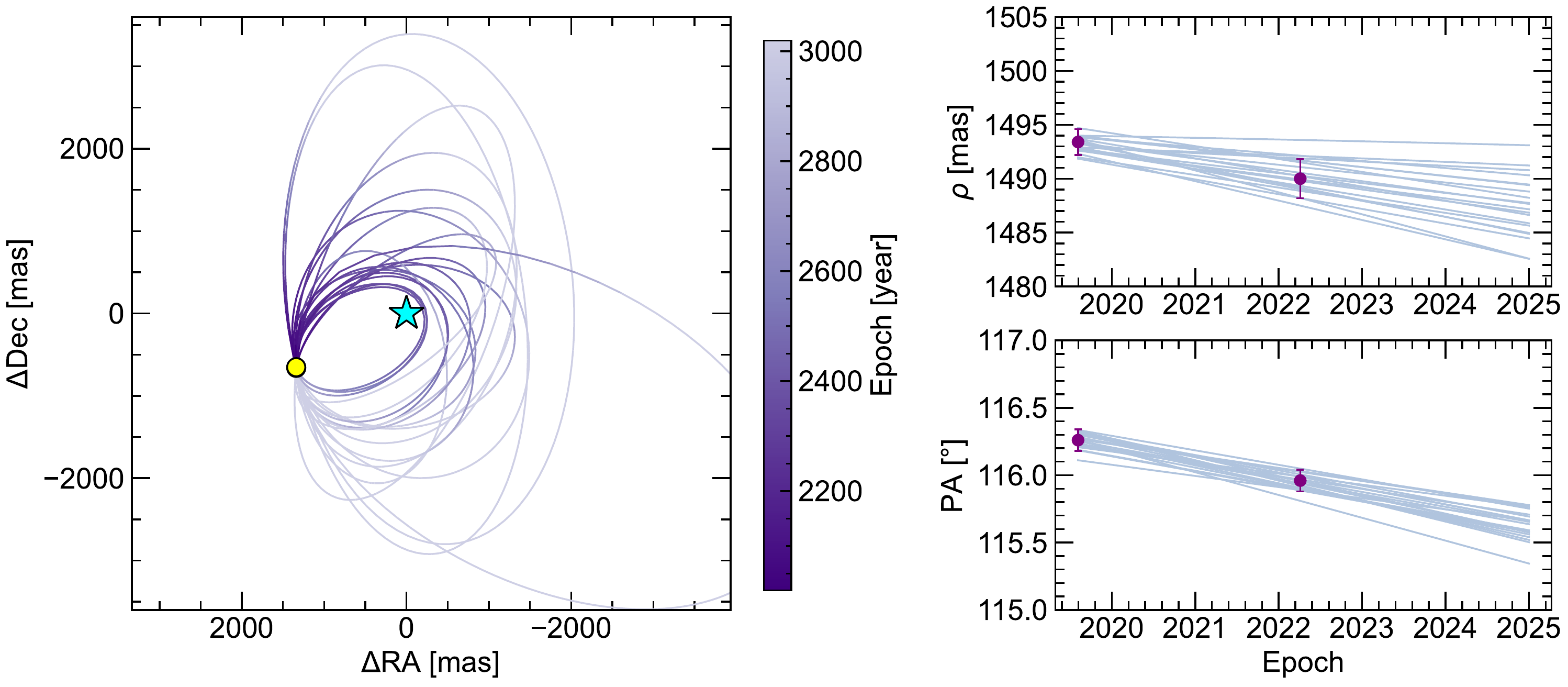} 
\caption{On the left: The motion of HIP 81208 C in RA and DEC over an entire orbital period for 20 random orbits drawn from the posterior distributions from the OFTI run. The change in the position of the companion over the two epochs is very small to be visible as separate data points as in the previous figure, and are thus represented by a single yellow solid circle. On the right: The projected separation $\rho$ and position angle PA (deg) of C over time, as  predicted from these 20 orbits. The symbols and color bar hold the same meaning as in the previous figure.}
\label{fig11}
\end{figure*}

Fig. \ref{fig10} and \ref{fig11} show 20 random orbits drawn from the posterior distributions from the OFTI run for B and C respectively. The left panels show the orbital motion of the respective companions in RA and DEC over an entire period of these orbits and the right panels show the projected separation $\rho~(mas)$ and position angle PA~(deg) predicted from these orbits over time. Also shown in the plots are the observed $\rho$ and PA of the companions in the two epochs along with error bars.

\begin{table*}[]
\centering
\resizebox{1.35\columnwidth}{!}{%
\begin{tabular}{l|l|l}
\hline\hline
& HIP 81208B & HIP 81208C \\ \hline
$K1-K2$ (mag) & $0.250\pm0.057$ & $0.252\pm0.064$ \\
\teff~(K) from \textsc{madys} & $2895^{+45}_{-40}$ & $3165^{+40}_{-60}$ \\
\teff~(K) from spectral fit & 2900 & \textemdash \\
Spectral type & M5 & tentatively M4 \\
Mass, \textit{M} ($M_{\odot}$) & $0.064^{+0.006}_{-0.007}$ &
$0.135^{+0.010}_{-0.013}$ \\
Orbital semi-major axis, \textit{a} (au) & $53.98^{+32.22}_{-15.00}$ & $234.27^{+168.65}_{-68.96}$ \\ 
Implied orbital period, \textit{P} (years) & $244.12\pm160.19$ & $2178.11\pm1657.05$ \\ 
Orbital inclination, \textit{i} ($^{\circ}$) & $46.61^{+15.71}_{-19.47}$ & $128.16^{+19.47}_{-15.36}$ \\ 
Orbital eccentricity, \textit{e} & $0.33^{+0.26}_{-0.22}$ & $0.38^{+0.29}_{-0.26}$ \\ \hline
\end{tabular}
}
\caption{Properties of companions HIP 81208B and C determined in this work using astrometric and photometric information from BEAST observations in the two epochs.}
\label{tab3}
\end{table*}

\subsection{Dynamical stability of the orbits}
\label{sec:dynamics}

\begin{figure*}
    \centering
    \includegraphics[width=\linewidth]{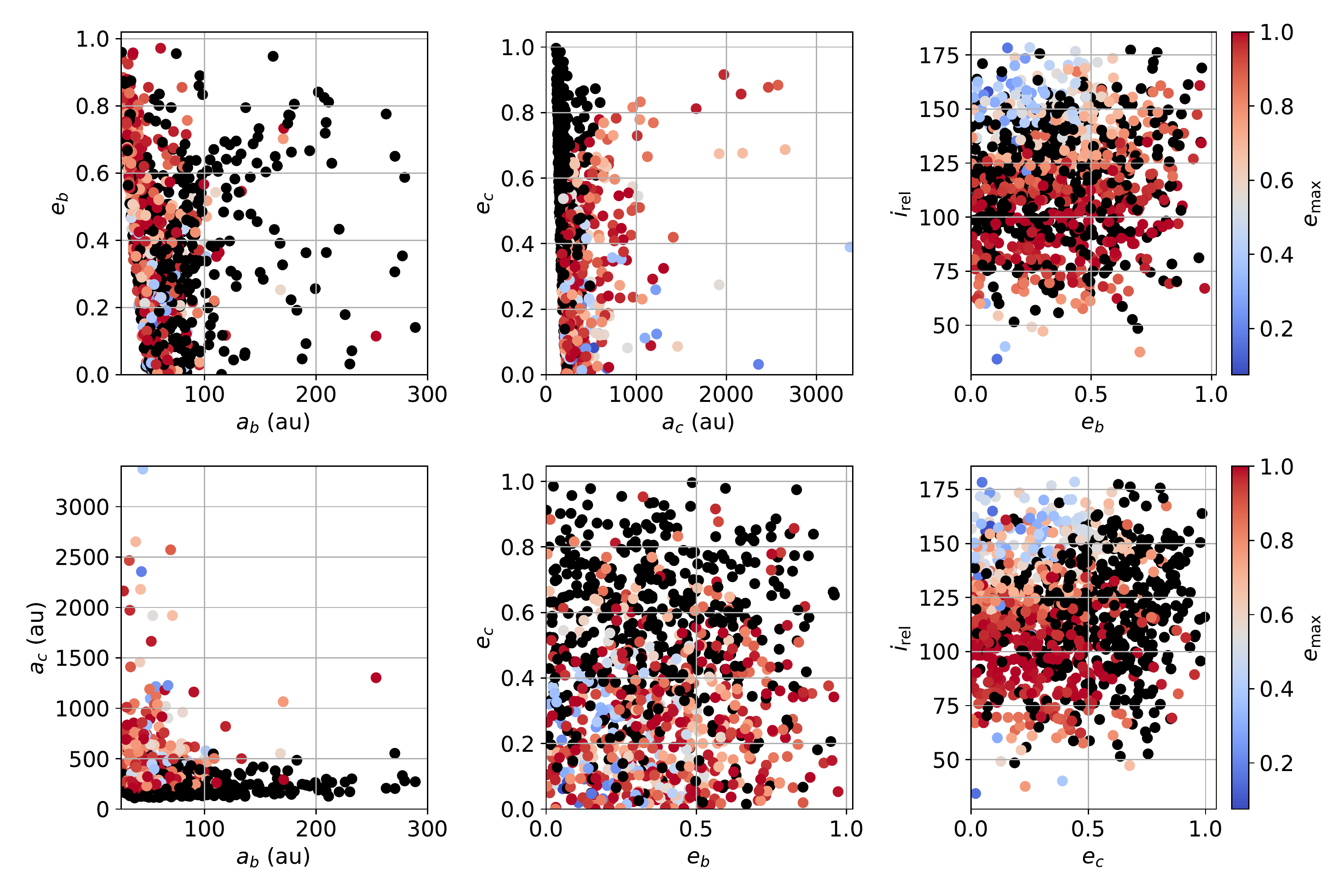}
    \caption{Maximum eccentricity reached by either orbit within $10^7$ years of evolution depending on the initial orbital elements. Black solid circles depict unstable cases ($e_{\rm max} > 1$, 44\% of the cases). These results were obtained by running $N$-body simulations starting from $1000$ solutions of the orbital fit. Configurations where orbits are crossing or almost crossing are always unstable. Orbits with relative inclination between $40$ and $130$ degrees are locked in a Kozai resonance, and may experience high eccentricity variations without compromising the system's stability.}
    \label{fig:maxe}
\end{figure*}

We can refine the current orbital constraints (depicted in Fig. \ref{fig9}) by ensuring that the configuration is stable for a timescale comparable to the age of the system ($\sim 10^7$ yr). Indeed, it is unlikely that we observe this system right before the ejection of one of its components. The stability of a triple system with non-negligible mass ratios requires a well separated hierarchical structure, with sufficient distance between the orbits. Moreover, if the eccentricities and/or relative inclination are high, then additional secular perturbations can destabilise orbits that were initially well separated. To evaluate the additional constraints ensuing from dynamical stability, we randomly pick 1000 couples of solutions to the orbital fits, and run $N$-body simulations of their future $10^7$ years of evolution. We use the IAS15 integrator in the Rebound package \citep{rein2015}. We adopt the following values for the masses: $M_* = 2.58~M_\odot$, $M_B = 0.064~M_\odot$, and $M_C = 0.135~M_\odot$. The maximum value of the eccentricity reached by either orbit is displayed on Fig.~\ref{fig:maxe} as a function of the semi-major axes, eccentricities and relative inclination. If this value is more than 1, it denotes instability (corresponding orbits are indicated as black solid circles in the figure). The results suggest that a large fraction of the orbital solutions (nearly half of them) are actually unstable. In particular, for the companion C, the lower right panel of Fig.~\ref{fig:maxe} shows that instability develops starting from $e_c\simeq0.5$, whatever the relative inclination (the chaos becomes too strong for the system to be locked in a Kozai regime). From the middle top panel of the figure, we also see that only wide separations ($a_c>500$ au) could help maintain the stability for $e_c>0.5$, however such separations are outside the interval of confidence for C (see Table \ref{tab3}). Hence, we can say that large eccentricities ($> 0.5$) are disfavored for C.

In the simulations, we notice that a significant proportion of solutions reach high eccentricities without becoming unstable. This is characteristic of the Von Zeipel-Lidov-Kozai \citep[ZLK][]{vonziepel1910,lidov1962,kozai1962} resonance, which occurs when the relative inclination is more than $\simeq 40$ degrees and less than $\simeq 130$ degrees. Looking at the whole set of solutions, the likelihood that this system is in a configuration of ZLK resonance is around 65\% (see Fig.~\ref{fig:inc}). The period of Kozai eccentricity oscillations for a triple system can be estimated approximately as \citep{takeda2005, ford2000}:
\begin{equation}
P_{Koz} \simeq P_1 \left( \frac{m_0+m_1}{m_2} \right) \left(\frac{a_2}{a_3}\right)^{3}\left(1-e_2^2\right)^{3/2},
\end{equation}
where \textit{P}, \textit{a}, \textit{e} are the orbital period, semi-major axis and eccentricity respectively and the indices 0, 1, 2 are, in this case, representative of the primary star HIP 81208, the brown dwarf companion B and the stellar companion C respectively. For the system HIP 81208, using the derived mass and orbital parameters in Sections \ref{sec:mass}, \ref{sec:orbit}, the Kozai oscillation period is $\simeq$0.3 Myr, which is much shorter compared to 17 Myr, the age of the system. So the system has enough time to have undergone atleast a few cycles of Kozai oscillations.   
Such systems are rare and precious, for ZLK resonances are at the core of many astrophysical problems such as hot Jupiters \citep[see][]{wu2003} or black hole mergers \citep[see][]{liu2019}. More generally, the relative inclination between the different orbits of a given system is far from being well constrained, and could bring important clues to the formation processes. In this regard, HIP 81208 probes an uncharted regime, between the planetary \citep{dupuy2022} and stellar masses \citep{tokovinin2017}, where it is unclear whether coplanarity or random inclinations should be favored.

\subsection{A potential `D' component}
\label{sec:potential_D}
In addition to the two companions reported in this study, the star Gaia DR3 6020420074469092608 (2MASS J16360769-3543514, WISEA J163607.70-354351) at a separation of 656.28\arcsec~(0.1823$^{\circ}$) and position angle of 92.046$^{\circ}$ is potentially noteworthy. The star is a faint \citep[G=$14.975\pm0.003$ mag;][]{gaia2020}, X-ray emitting source detected by the Rosat All-Sky Survey \citep[1eRASS J163607.7-354351, 2RXS J163607.8-354352; see][]{Boller2016}. Its parallax, $6.6742\pm0.0297$ mas \citep{gaia2020}, differs from that of HIP 81208 only by $\sim3\sigma$ and the proper motion \citep[$\mu_{\alpha*}=-9.622\pm0.035$ \masyr, $\mu_{\delta}=-24.695\pm0.029$ \masyr;][]{gaia2020} only by $\sim1.12$ \masyr ($856\pm56$ ms$^{-1}$). Given the total mass of the HIP 81208 triple system to be $M_{tot}=2.779~M_{\odot}$, the tidal radius of HIP 81208 is then 1.35 pc$\times(M_{tot}/M_{\odot})^{1/3}$=1.898 pc \citep{mamajek2013, jiang2010}. At a distance of 148.7 pc, this translates to a projected tidal radius of (180/$\pi$)$\times$(1.898 pc/148.7 pc)=0.7313$^{\circ}$, or 2632.75\arcsec. Since the star is within 656.28\arcsec of HIP 81208, it is possible that it could be within the tidal influence of our target. This fact, along with the similar parallax and proper motion to HIP 81208, suggest a possibility that this star could be a potential `D' companion in this currently-triple system. However, we cannot reliably assess this possibility at the moment since not much is known about this star. Future observations of Gaia DR3 6020420074469092608 with instruments like GRAVITY \citep{gravity2017} could help provide more information on its projected 2D motion in the sky. In addition, radial velocity information from future spectroscopic observations can also help determine its 3D motion. If the star shares the same 3D motion in the sky as HIP 81208, this could be a strong indication that this star is bound to our target.

\section{Conclusions}
\label{s:summary}
In this work, we describe our observations of the B9V star HIP 81208 using the SPHERE instrument at VLT as part of the BEAST survey, in two different epochs, obtaining both IFS $YJH$ spectroscopy as well as IRDIS $K1$, $K2$ imaging data for the target. 
We report the discovery of two lower mass companions to the star; an inner companion HIP 81208B at $0.325\pm0.001$\arcsec and an outer companion HIP 81208C at $1.492\pm0.001$\arcsec projected separation. Using its spectrum over the $YJH-K1-K2$ wavelength range, we determine that HIP 81208B is of spectral type M5. Analysing the obtained photometry, we estimate the masses of HIP 81208B and C to be $67^{+6}_{-7}~M_{J}$ and $0.135^{+0.010}_{-0.013}~M_{\odot}$ respectively, indicating B to be most likely a brown dwarf and C to be a low-mass star. Both companions are confirmed to be physically bound to the primary by means of proper motion analysis of the background stars in the data. 

Using astrometric information over the two epochs, we were able to constrain the orbits of B and C to the most probable semi-major axis values of $53.98^{+32.22}_{-15.00}$ au and $234.27^{+168.65}_{-68.96}$ au respectively. The relative inclination between the two orbits is high, making them appear to orbit in an opposite sense to each other. Table \ref{tab3} summarises all main parameters we derived for the two companions in this work. The orbital solutions we derived for the companions indicate that this system is likely to be in a Kozai resonance, in which orbits with high relative inclinations can remain stable despite reaching high values of eccentricities. This makes HIP 81208 a system to be subjected to careful further study, to understand more about orbital dynamics in systems where the companions are in between planetary and stellar mass regimes. 

Our observations have, thus, presented strong evidence that HIP 81208 is a triple system, with two low-mass companions in a very interesting configuration around the star. If the nearby star Gaia DR3 6020420074469092608 turns out to be physically bound to HIP 81208, the system may even be quadruple. Further observations are thus useful for HIP 81208, both to investigate the possibility of perhaps a `D' component in the now known triple system, and to understand more about the orbital dynamics in the system. The orbital parameters derived in this study for HIP 81208 B and C are not tightly constrained as can be seen from their large $\pm1\sigma$ errors, since they are based on a small time baseline. Future BEAST or GRAVITY epochs will aid in providing longer time baseline astrometry for both the companions, which could help establish more stringent constraints on the orbital parameters for this very interesting system.

\begin{acknowledgements}
M.J. gratefully acknowledges funding from the Knut and Alice Wallenberg Foundation. AV and PD acknowledges funding from the European Research Council (ERC) under the European Union’s Horizon 2020 research and innovation programme (grant agreement No. 757561, COBREX; grant agreement n° 885593).  This publication makes use of CDS and NASA/ADS services, as well as VOSA, developed under the Spanish Virtual Observatory (https://svo.cab.inta-csic.es) project funded by MCIN/AEI/10.13039/501100011033/ through grant PID2020-112949GB-I00. 
VOSA has been partially updated by using funding from the European Union's Horizon 2020 Research and Innovation Programme, under Grant Agreement no. 776403 (EXOPLANETS-A). This work has also made use of the SPHERE Data Centre, jointly operated by OSUG/IPAG (Grenoble), PYTHEAS/LAM/CeSAM (Marseille), OCA/Lagrange (Nice), Observatoire de Paris/LESIA (Paris), and Observatoire de Lyon (OSUL/CRAL). This research has made use of the SIMBAD database and VizieR catalogue access tool, operated at CDS, Strasbourg, France. This work has made use of data from the European Space Agency (ESA) mission {\it Gaia} (\url{https://www.cosmos.esa.int/gaia}), processed by the {\it Gaia} Data Processing and Analysis Consortium (DPAC,
\url{https://www.cosmos.esa.int/web/gaia/dpac/consortium}). Funding for the DPAC has been provided by national institutions, in particular the institutions participating in the {\it Gaia} Multilateral Agreement.This work is supported by the French National Research Agency in the framework of the Investissements d’Avenir program (ANR-15-IDEX-02), through the funding of the "Origin of Life" project of the Univ. Grenoble-Alpes. This work is also supported by the PRIN-INAF 2019 "Planetary systems at young ages (PLATEA)". Part of this research was carried out at the Jet Propulsion Laboratory, California Institute of Technology, under a contract with the National Aeronautics and Space Administration (80NM0018D0004).  
\end{acknowledgements}

\begin{appendix}

\section{Properties of IRDIS sources}
\begin{table}[!htbp]
\centering
\caption{Astrometric and photometric properties of all the sources detected in the IRDIS FoV in both epochs.}
\label{table:cc_info}
\centering
\scalebox{0.9}{
\[
\begin{array}{c|cccc|cccc}
\hline \hline
\multicolumn{1}{c}{} & \multicolumn{4}{c}{\text{First epoch}} & \multicolumn{4}{c}{\text{Second epoch}} \\
\hline
\multicolumn{9}{c}{\text{Comoving companions}} \\
\hline
\text{ID} & d \text{ (mas)} & \text{PA} (^\circ) & \Delta K1 \text{ (mag)} & \Delta K2 \text{ (mag)} & d \text{ (mas)} & \text{PA} (^\circ) & \Delta K1 \text{(mag)} & \Delta K2 \text{(mag)} \\
\hline        
0 & 320.9\pm 1.0	& 356.55 \pm 1.72 & 6.88 \pm 0.05 & 6.64 \pm 0.07 & 328.7 \pm 1.0 & 0.43 \pm 0.13 & 6.85 \pm 0.12 & 6.59 \pm 0.12 \\
1 & 1493.4 \pm 1.2 & 116.26 \pm 0.08 & 5.83 \pm 0.05 & 5.59 \pm 0.07 & 1490.0 \pm 1.8 & 115.96 \pm 0.08 & 5.77 \pm 0.12 & 5.51 \pm 0.12 \\
\hline        
\multicolumn{9}{c}{\text{Background sources}} \\
\hline
2 & \text{\textemdash} & \text{\textemdash} & \text{\textemdash} & \text{\textemdash} & 2505 \pm 13 & 91.15 \pm 0.24 & 14.23 \pm 0.24 & 14.13 \pm 0.75 \\
3 & 3291 \pm 4 & 245.01 \pm 0.08 & 12.86 \pm 0.07 & 12.64 \pm 0.21 & 3261 \pm 6 & 245.93 \pm 0.09 & 12.85 \pm 0.13 & 12.74 \pm 0.17 \\
4 & 4515 \pm 5 & 116.10 \pm 0.08 & 12.91 \pm 0.07 & 12.62 \pm 0.17 & 4492 \pm 8 &	115.44 \pm 0.08 & 12.90 \pm 0.13 & 12.75 \pm 0.21 \\
5 & 4723 \pm 8 & 254.76 \pm 0.09 & 13.40 \pm 0.10 & 13.52 \pm 0.21 & 4718 \pm 10 &	255.38 \pm 0.10 & 13.43 \pm 0.15 & 13.36 \pm 0.45 \\
6 & 4822 \pm 14 & 276.86 \pm 0.13 & 14.27 \pm 0.10 & \text{\textemdash} & \text{\textemdash} &	\text{\textemdash} & \text{\textemdash} & \text{\textemdash} \\
7 & 4982 \pm 19 & 120.11 \pm 0.20 & 14.37 \pm 0.18 & \text{\textemdash} & \text{\textemdash} &	\text{\textemdash} & \text{\textemdash} & \text{\textemdash} \\
8 & 5460 \pm 12 & 295.13 \pm 0.11 & 14.01 \pm 0.13 & \text{\textemdash} & 5494 \pm 15 &	295.70 \pm 0.14 & 14.23 \pm 0.20 & 14.15 \pm 0.71 \\
9 & 5807 \pm 18 & 267.26 \pm 0.14 & 14.33 \pm 0.26 & 14.67 \pm 0.41 & \text{\textemdash} &	\text{\textemdash} & \text{\textemdash} & \text{\textemdash} \\
10 & 6382 \pm 11 & 328.85 \pm 0.15 & 13.06 \pm 0.14 & 13.71 \pm 0.47 & 6434 \pm 14 &	329.08 \pm 0.16 & 13.22 \pm 0.20 & \text{\textemdash} \\
11 & 6436 \pm 5 & 313.78 \pm 0.08 & 11.07 \pm 0.10 & 10.91 \pm 0.10 & 6467 \pm 10 &	314.13 \pm 0.09	& 11.31 \pm 0.17 & 11.15 \pm 0.15 \\
12 & \text{\textemdash} & \text{\textemdash} & \text{\textemdash} & \text{\textemdash} & 6645 \pm 9 &	179.59 \pm 5.34 & 10.48 \pm 0.16 & 10.45 \pm 0.16 \\
\hline        
\end{array}
\]
}
\end{table}

\end{appendix}
\end{document}